%% file: FINAL_VERSION.tex
\documentclass[journal]{IEEEtran}
\normalsize

\usepackage{url}

\usepackage[T1]{fontenc}
\usepackage[dvips]{graphicx}
\usepackage{array}    
\usepackage{amssymb,amsthm, mathtools}%
\usepackage{amsmath}
\usepackage{float}
\usepackage{subfig}
\usepackage[english]{babel}
\usepackage{import}
\usepackage{xcolor}
\usepackage{tikz}
\usepackage{pgfplots}
\usepackage{cite}

\usepackage{tabularx}
\usepackage{multicol}

\usepackage{algorithm}
\usepackage{algpseudocode}

\algsetblockdefx{beginIni}{endIni}{1}{}
{\textbf{Initialize:}}%
{\textbf{Initialize}}

\algsetblockdefx{beginChEst}{endChEst}{2}{}
[2]{#1 \textbf{channel estimation} #2}
[2]{#1 \textbf{channel estimation} #2}

\algsetblockdefx{beginChEstData}{endChEstData}{2}{}
[2]{#1 \textbf{channel estimation} #2}
[2]{#1 \textbf{channel estimation} #2}

\algsetblockdefx{beginLinComb}{endLinComb}{2}{}
{\textbf{Linear combining}}
{\textbf{Linear combining}}

\algsetblockdefx{beginDataDec}{endDataDec}{3}{}
[1]{\textbf{Data decoding} #1}
{\textbf{Data decoding}}

\usetikzlibrary{shapes.multipart}
\usetikzlibrary{patterns,calc,fit,arrows,chains,positioning}

\newdimen{\Offset}

\pgfplotsset{
	width=0.62\textwidth,
	height=0.45\textwidth,
}

\pgfplotsset{xindexrange/.style 2 args={
		x filter/.code={
			\ifnum\coordindex<#1\fi
			\ifnum\coordindex>#2\fi
		}
	}}

\usepackage{picture}
\graphicspath{{./Figures/}}

\newtheoremstyle{mytheoremstyle} 
{\topsep}                    
{0pt}                    
{\itshape}                   
{}                           
{\bfseries}                   
{.}                          
{.5em}                       
{}  

\theoremstyle{mytheoremstyle}

\newtheorem{theorem}{Theorem}
\newtheorem{lemma}{Lemma}

\newtheorem{remark}{Remark}
\newtheorem{corollary}{Corollary}

\newcommand{\bb}[1]{\mathbb{#1}}
\renewcommand{\bf}[1]{\mathbf{#1}}
\renewcommand{\rm}[1]{\mathrm{#1}}
\renewcommand{\cal}[1]{\mathcal{#1}}
\newcommand{\bfs}[1]{\boldsymbol{#1}}
\renewcommand{\sf}[1]{\mathsf{#1}}

\def\Bw{B_\textsc{w}}

\def\ep{\epsilon}

\def\qul{\varrho} 
\def\pul{\rho}

\def\Tp{\tau_p}
\def\Tc{\tau_c}

\DeclareMathOperator*{\minimize}{\rm{minimize}}
\DeclareMathOperator*{\maximize}{\rm{maximize}}
\DeclareMathOperator*{\st}{\rm{subject\;to}}

\newcolumntype{L}[1]{>{\raggedright\let\newline\\\arraybackslash\hspace{0pt}}m{#1}}
\newcolumntype{C}[1]{>{\centering\let\newline\\\arraybackslash\hspace{0pt}}m{#1}}
\newcolumntype{R}[1]{>{\raggedleft\let\newline\\\arraybackslash\hspace{0pt}}m{#1}}

\makeatletter
\newcommand{\manuallabel}[2]{\def\@currentlabel{#2}\label{#1}}
\makeatother
\allowdisplaybreaks

\begin{document}
%
\title{Optimal Per-Antenna ADC Bit Allocation in Correlated and Cell-Free Massive MIMO}

\author{Daniel~Verenzuela,~\IEEEmembership{Student~Member,~IEEE,}~Emil~Bj\"ornson, \IEEEmembership{Senior Member, IEEE,}~and~Michail~Matthaiou,~\IEEEmembership{Senior~Member,~IEEE}
\thanks{\copyright 2021 IEEE. Personal use of this material is permitted. Permission from IEEE must be obtained for all other uses, in any current or future media, including reprinting/republishing this material for advertising or promotional purposes, creating new collective works, for resale or redistribution to servers or lists, or reuse of any copyrighted component of this work in other works.}
\thanks{D. Verenzuela and E. Bj\"{o}rnson are with the Department
	of Electrical Engineering (ISY), Link\"{o}ping University, Link\"{o}ping, SE-58183 Sweden. In addition, E. Bj\"{o}rnson is with the Department of Computer Science, KTH Royal Institute of Technology, 100 44 Stockholm, Sweden (e-mail: daniel.verenzuela@liu.se; emilbjo@kth.se).}
\thanks{M.  Matthaiou  is  with  the  Institute  of  Electronics,  Communications  and Information  Technology  (ECIT),  Queen's  University  Belfast,  Belfast,  BT3 9DT, U.K. (e-mail: m.matthaiou@qub.ac.uk).}
\thanks{This paper has received funding from ELLIIT and the Swedish Foundation for Strategic Research (SSF). The work of M. Matthaiou was supported by the EPSRC, U.K., under Grant EP/P000673/1 and by a research grant from the Department for the Economy Northern Ireland under the US-Ireland R\&D Partnership Programme. }
}

\maketitle

\begin{abstract}

In Massive MIMO base stations (BSs), the hardware design needs to balance high spectral efficiency (SE) with low complexity. The level of hardware impairments (HWIs) indicates how strong the signal distortion introduced by hardware imperfections is. In particular, the analog-to-digital converters (ADCs) have an important impact on signal distortion and power consumption. This article addresses the fundamental problem of selecting the optimal hardware quality in the Massive MIMO space. In particular, we examine the optimal HWI and ADC bit allocation per BS antenna to maximize the SE. The results show that in co-located arrays with low channel gain variations across antennas, equal ADC bit allocation is optimal. In contrast, cell-free Massive MIMO systems benefit the most from optimizing the ADC bit allocation achieving improvements in the order of 2 [bit-per-channel-use] per user equipment when using regularized zero-forcing (RZF). In addition, when including the impact of power consumption in cell-free Massive MIMO with RZF, allocating low values of mixed ADC bit resolutions across the BS antennas can increase the energy efficiency up to 30\% compared to equal ADC bit allocation.
\end{abstract}

\IEEEpeerreviewmaketitle

\section{Introduction}
\label{sec:introduction}

In this information age, the demands for data traffic rise every year and wireless communications are essential to support the increase in mobile connectivity \cite{cisco_GMDT_2019}. In parallel, the power consumption (PwC) of information and communication technologies (ICT) is rising and becoming an important economical and environmental burden \cite{Co2_footprint,ICT_electricity_2012}. Thus, to satisfy the	high data traffic demands and maintain or even reduce the PwC of ICT, it is imperative to make improvements in spectral efficiency (SE) and energy efficiency (EE) of wireless networks.

In the development and implementation of 5G networks,  Massive MIMO is a key multiple antenna technology that provides both high SE and EE. This is realized by using a large number of antennas at the base station (BS) to serve several user equipments (UEs) on the same time-frequency resource with spatial multiplexing techniques \cite{16_Marzetta_MAMIMO_book, Bjornson2017_MaMIMObook,20_zhang_tech5G}.  

In 3G and 4G cellular networks, the BSs account for the largest portion of the total PwC \cite{Green_radio_2011}, and within macro BSs, the power amplifiers and site cooling are responsible for over 80\% of their consumption\cite{Green_cellNet_2011}. By directing signals into narrow beams using a large number of BS antennas, Massive MIMO can substantially reduce the transmission PwC in comparison to current macro BSs \cite{16_Marzetta_MAMIMO_book, Bjornson2017_MaMIMObook}. As a result, power amplifiers would require much less power to operate, and in turn, the cooling requirements would also be reduced since the power amplifiers are predominantly responsible for heat dissipation. From another perspective, Massive MIMO can be used to serve more UEs while maintaining the same PwC as compared to current cellular networks, which in turn, increases EE. However, increasing the number of BS antennas requires more hardware components that will analogously increase the PwC. The above discussion reveals a critical component of Massive MIMO operation, that is the hardware design. We will now overview some recent advances that have reported in the related literature. 

We first recall, that it has been shown that Massive MIMO BSs can operate with low-end hardware that reduces PwC while still supporting high SE \cite{Emil_EE_hardware, Emil_scaling_laws, Impact_MAMIMO_hard}. An important component in the receiver hardware of BS antennas is the analog-to-digital converter (ADC) for its impact on signal distortion and PwC\cite{ADC_p_mod_A3}. Each antenna element requires two ADCs (for in-phase and quadrature chains), which means hundreds or more ADCs would be needed in a fully digital Massive MIMO BS. A key feature of Massive MIMO is the ability to coherently combine signals in the same time scale as the small-scale fading variations. This means that the digital signals from the ADCs need to be jointly processed in real-time. Thus, the ADCs can not only increase the PwC, but can also create a bottleneck in the fronthaul processing. To manage these effects, low ADC bit resolution has been proposed for Massive MIMO BS antennas, which can, in theory, support high performance in terms of bit-error-rate, SE, and PwC \cite{low_ADC_desset,S_Jacobsson_Durisi_ADCs_UL,MollenCLH16a_ach_rate_ADCs,Verenzuela_2017_HWI_ADC}.

In the aforementioned works, low and equal ADC bit resolution is assumed across BS antennas. Alternatively, a mixed-ADC approach has been proposed in \cite{N_Liang_MixedADC, Zhang2016_Mixed_ADC}, where the ADC bit resolution is allowed to be different across BS antennas and vary at high speed following the small-scale fading variations. In this context, \cite{Ahmed2018_PerfectCSI,Zhang2016_Mixed_ADC} studied the optimal ADC bit allocation to maximize the SE in millimeter wave and Massive MIMO systems, respectively. However, these works do not consider the effect of imperfect channel state information (CSI) which is crucial to accurately evaluate the performance of Massive MIMO. Recently, \cite{Pirzadeh2018_Mixed_ADC, Yuan_mxADCIoT_MM2019} studied a mixed-ADC system under imperfect CSI, where a small portion of the BS antennas have high-resolution ADCs and the remaining have 1-bit ADCs \cite{Pirzadeh2018_Mixed_ADC}, and low-resolution ADCs \cite{Yuan_mxADCIoT_MM2019}.  In \cite{Yuan_mxADCIoT_MM2019}, the aim was to optimize the access phase of internet-of-things devices considering the impact of mixed-ADCs in a cell-free Massive MIMO system. The authors concluded that by optimizing the UE access to the network, high gains in sum SE and EE can be achieved. In \cite{Pirzadeh2018_Mixed_ADC}, the authors considered uncorrelated Rayleigh fading and concluded that mixed-ADCs architectures are most beneficial for high signal-to-noise ratio (SNR) while using zero-forcing processing.  In our previous work  \cite{17_Verenzuela_mixed_ADC}, majorization theory was used to assess the impact of ADC bit allocation under a total ADC bit budget and uncorrelated Rayleigh fading. The results indicate that low and equal ADC bit allocation (e.g., 3-4 bits) is the optimal strategy to maximize the SE and minimize PwC.

	\subsection{Contributions}
	\label{subsec:contribution}
This article studies the uplink of a Massive MIMO system, where the ADC bit resolution can be customized based on the large-scale fading (LSF) variations of the channel. These variations occur due to the movement of UEs and obstacles in a macroscopic scale\footnote{This scale can span orders of magnitude larger than the wavelength of propagating signals.} or by serving different UEs in the coverage area over time. The main contributions of this paper are given as follows:
\begin{itemize}
	\item To the best of our knowledge, this is the first article that optimizes the ADC bit allocation based on a rigorous achievable SE analysis with \textit{spatially correlated channels} and {LSF variations} across the BS antennas under imperfect CSI. Our analysis considers a co-located BS antenna array and a cell-free Massive MIMO deployment, where the BS antennas are distributed over the service area. In practice, the channels between different BS antennas are indeed spatially correlated and exhibit LSF variations \cite{FTufvesson15_measured_MAMIMO}. These channel phenomena affect the dynamic range of the received signal power across the BS antennas, and in turn, the quantization distortions. This makes the ADC bit allocation fundamentally different from scenarios with uncorrelated fading, as in \cite{Pirzadeh2018_Mixed_ADC,17_Verenzuela_mixed_ADC}. In addition, the LSF variations occur over a much slower time scale compared to the small scale fading changes.\footnote{ The LSF fading changes in seconds or hundreds of milliseconds, whereas the small-scale fading changes within a few milliseconds or microseconds \cite{Bjornson2017_MaMIMObook}.} This simplifies the mixed-ADC implementation and provides the necessary time to conduct further optimization of the ADC bit allocation, in contrast to the prior work in \cite{N_Liang_MixedADC,Ahmed2018_PerfectCSI,Zhang2016_Mixed_ADC,Pirzadeh2018_Mixed_ADC}, where high-speed switches are required to track the small-scale fading.
	
	\item A general optimization framework is proposed to allocate the ADC bits across the BS antennas. A detailed analysis of the minimum pilot distortion and maximum product of SINRs ADC bit allocation is given under a sum of ADC bits (ADC bit budget) or a PwC constraint. The ADC bit budget constraint allows for a direct evaluation of the allocation of ADC bits per BS antenna rather than focusing on the total number of ADC bits available. On the other hand, the PwC constraint accounts for a more practical scenario compared to the ADC bit budget, limiting the aggregated power consumed by all ADCs. The impact of optimal ADC bit allocation is evaluated in terms of SE and EE.
	   
	\item  A closed-form optimal ADC bit allocation strategy is proposed to minimize the pilot distortion under an ADC bit budget. This strategy is simple to implement and it reveals important insights into the relationship between the ADC bit allocation and the received power level at each BS antenna. 
	
	\item A closed-form achievable SE with maximum ratio (MR) combining is derived as a rigorous lower bound on the capacity. This SE expression is used to formulate geometric programs to allocate the ADC bits across the BS antennas.

\end{itemize}

The rest of this article is organized as follows: Section~\ref{sec:sys_model} describes the system model. Section~\ref{sec:Ch_est} explains the channel estimation procedure, the pilot distortion minimization problem along with its solution in closed-form, and introduces the main study cases for the numerical results. To evaluate the SE performance, an achievable closed-form SE expression is derived in Section~\ref{sec:AchSE}. In Section~\ref{sec:Op_HWI_allc}, optimal ADC bit allocation strategies are developed based on the aforementioned SE expression. Section~\ref{sec:PwC} studies the effect of PwC on the ADC bit allocation and evaluates the performance in terms of EE. Finally, Section~\ref{sec:conclusion} concludes our work.

\textbf{Notation:} The transpose, conjugate, conjugate transpose, and inverse are given by $(\cdot)^T$, $(\cdot)^*$, $(\cdot)^H$, and $(\cdot)^{-1}$ respectively. The set of natural and complex numbers are denoted as $\bb{N}$ and $\bb{C}$, respectively. Matrices and vectors are represented by upper and lower bold case letters, as $\bf{X}$ and $\bf{x}$, respectively. The element in the $i^{th}$ row and $j^{th}$ column of a matrix $\bf{X}$ is given by $[\bf{X}]_{ij}$ while $[\bf{x}]_{i}$ denotes the $i^{th}$ element of $\bf{x}$. The $i^{th}$ column of $\bf{X}$ is denoted by $\bf{x}_i$. The notation $\rm{diag}(\bf{X})$, corresponds to the matrix $\bf{X}$ with all non-diagonal elements set to zero, whereas, $\rm{diag}(\bf{x})$ is a diagonal matrix with the elements of $\bf{x}$ in the diagonal. The absolute value is denoted as $|\cdot|$ and  $\rm{abs}(\bf{X})^2$ takes the squared absolute value on every element of $\bf{X}$.

\setlength{\textfloatsep}{8pt}

\section{System Model}
\label{sec:sys_model}
Consider the uplink of a single-cell Massive MIMO system, where the BS has $M$ antennas and serves $K$ single-antenna UEs via spatial multiplexing. The communication channel follows a block fading model in which the channel is considered static for a time period $T_c$ and frequency-flat within a bandwidth $B_c$. The total number of complex samples enclosed within the time $T_c$ and bandwidth $B_c$ is denoted as $\Tc = T_c B_c$, which in turn, forms a coherence block. The block fading model assumes that the channel realizations remain fixed within a given coherence block and change independently at random from one coherence block to another. The channel between the $M$ BS antennas and UE $k$, denoted as $\rm{UE}_k$, is defined as $\bf{h}_k \sim \cal{CN}(\bf{0}, \bf{R}_k)$. The spatial correlation matrix of the channel $\bf{h}_k$ is given by $\bf{R}_k$ and $\bar{\beta}_k = \rm{tr}(\bf{R}_k)/M$  is the average channel gain that corresponds to the LSF effect. The diagonal elements of $\bf{R}_k$ can be different which allows for modeling both co-located and cell-free Massive MIMO channels using the same system model. Special cases of these channel models are given in Section~\ref{subsec:HWIsim_pilot_dist}. The receiver hardware connected to each BS antenna is assumed to be affected by hardware impairments (HWIs), that are modeled as an additive distortion with an energy proportional to the signal energy with proportionality constant $\ep_m^2$ $\forall m \in \{1,\ldots,M\}$. Hence, $\ep_m = 0$ corresponds to perfect hardware at the $m^\text{th}$ BS antenna. The received signal at the BS for an arbitrary sample in a given coherence block is
\begin{equation}
	\bf{y} = \sum_{i = 1}^{K}  \bf{h}_i x_i + \bf{n} + \bf{e}
	\label{eq:rx_signal}
\end{equation} 
where $x_i$ is the transmitted signal from $\rm{UE}_i$ which can be composed of a data or pilot symbol.  
The thermal noise is defined as $\bf{n}\sim\cal{CN}(\bf{0},\sigma^2\bf{I}_M)$, where $\sigma^2$ is the average noise energy per symbol. By defining $\bf{D}_{\ep} = \rm{diag}(\ep_1,\ldots,\ep_M)$, the distortion caused by HWIs is modeled as
\begin{equation}
	\bf{e} = \bf{D}_{\ep} \left(\vphantom{\sum_{i}^{K}}\right.  \underbrace{\sum_{i = 1}^{K}\bb{E}\left\{\left|x_i\right|^2\right\}\rm{diag}\left( |[\bf{h}_i]_1|^2,\ldots, |[\bf{h}_i]_M|^2\right)}_{\text{=}\bf{D}_h} \left.\vphantom{\sum_{i}^{K}}\right)^{\frac{1}{2}} \bf{r} 
	\label{eq:distortion_e}
\end{equation}
where the randomness introduced by the hardware distortion is given by $\bf{r}\sim\cal{CN}(\bf{0},\bf{I}_M)$. 
The total amount of instantaneous energy\footnote{The term instantaneous energy is used to specify that the hardware distortion depends on the small-scale fading. Essentially, it corresponds to computing  the expected value of the received energy conditioned on the channel realizations.} received at the $m^{th}$ BS antenna is denoted as $[\bf{D}_h]_{mm}$. Since the pilot and data signals experience the same channel realizations, the term $\bf{D}_h$ allows for a clear evaluation of the hardware distortion within the channel estimation process and its impact on the SE.

\begin{remark}
	\label{rem:HWI_model}
	
	The distortion model in \eqref{eq:distortion_e} is based on Bussgang's decomposition, where the effect of a nonlinear deterministic operation is expressed as an additive distortion that is uncorrelated with the input. A recent review of this model is available in
	\cite{Demir2021a}.
	The model has been used to represent different types of physical HWIs that cause nonlinear distortions to the received signals \cite{S_Jacobsson_Durisi_ADCs_UL,Pirzadeh2018_Mixed_ADC,Bjornson2019_HWdist, Bjornson2017_MaMIMObook}. The quantization is such a nonlinear deterministic operation performed at the receiver and it depends on the received signal energy which is affected by the channel. Thus, the distortion model in \eqref{eq:distortion_e} is well suited for analyzing the distortions introduced by the ADCs.  
\end{remark}

The model for the hardware distortion in \eqref{eq:distortion_e} can be used to characterize the impact of HWIs on the SE of Massive MIMO systems, for any source of distortion that is proportional to the instantaneous received signal energy.\footnote{This means the the results for HWI allocation can be extended to design the quality of other hardware components apart from the ADCs.} In particular, the quantization distortion
\begin{align}
\ep_m &= \zeta_m 2^{-b_m}
\label{eq:ADC_Ce}	
\end{align}	
is used to map the ADC bit resolution and the level of HWI \cite{17_Verenzuela_mixed_ADC}. The term $\zeta_m$ depends on the saturation level of the ADC with practical values within $ 1<\zeta_m<2$, and $b_m$ is the ADC bit resolution at the $m^\text{th}$ BS antenna.

\begin{remark}
	In real systems, there is correlation between the distortion from different BS antenna elements and the model in \eqref{eq:distortion_e} neglects this effect. However, this correlation has been 
shown to have limited impact when many UEs are served \cite{Bjornson2019_HWdist}, particularly when it comes to coarse quantization, and it is a common practice to neglect it \cite{Pirzadeh2018_Mixed_ADC,S_Jacobsson_Durisi_ADCs_UL,MollenCLH16a_ach_rate_ADCs,Verenzuela_2017_HWI_ADC}. 
\end{remark}
Furthermore, the randomness of the distortion in \eqref{eq:distortion_e} is assumed to  be independent of the transmitted signal which is not true in practice but it allows for a tractable analysis that yields a closed-form expression for the SE under MR combining. Thus, in Section~\ref{sec:AchSE}, the SE with the distortion model in \eqref{eq:distortion_e} is compared against that of an exact quantization model to showcase the validity of SE expressions obtained with \eqref{eq:distortion_e} for optimizing the ADC bit allocation.

\section{Channel estimation}
\label{sec:Ch_est}
The channel is estimated at the BS based on pilot sequences transmitted by the UEs in the uplink. Thus, $\Tp$ out of $\Tc$ samples of the coherence block are reserved for channel estimation. The pilot transmitted by $\rm{UE}_k$ is denoted by $\bfs{\phi}_k \in \bb{C}^{\Tp \times 1}$, whose elements have unit modulus, that is, $|[\bfs{\phi}_k]_j| = 1$ $\forall j \in \{1,\ldots,\Tp\}$. The pilots are assumed to be mutually orthogonal and each UE transmits a different pilot, such that, $\bfs{\phi}_i^H \bfs{\phi}_k = 0$ if $i \neq k$ and  $\bfs{\phi}_i^H \bfs{\phi}_k = \Tp$ if $i = k$. The transmitted pilot signal from $\rm{UE}_k$ is defined by the vector $\bf{x}_k = \sqrt{\qul_k} \bfs{\phi}_k$ corresponding to $\Tp$ instances of $x_k$ in \eqref{eq:rx_signal}, thus, the received pilot signal is 
\begin{equation}
	\bf{Y}_\textsc{p} = \sum_{i = 1}^{K}  \bf{h}_i \sqrt{\qul_i} \bfs{\phi}_i^T  + \bf{N} + \bfs{\Xi} \in \bb{C}^{M \times \Tp}
\end{equation}
where $\qul_i$ is the energy per pilot symbol transmitted by $\rm{UE}_i$ such that $| [\bf{x}_i]_j |^2 = \qul_i$ $\forall j\in \{1,\ldots,\Tp\}$. The matrix $\bf{N}$ with i.i.d. elements defined as ${[\bf{N}]_{mj}\sim\cal{CN}(0,\sigma^2)}$ represents the thermal noise, and  ${\bfs{\Xi} = [\bf{e}_1,\ldots,\bf{e}_{\Tp}]}$ corresponds to the hardware distortion matrix. The hardware distortion is defined as in the previous section, such that $\bf{e}_j = \bf{D}_{\ep} \bf{D}_h^{\frac{1}{2}} \bf{r}_j$, where $\bf{r}_j\sim\cal{CN}(\bf{0},\bf{I}_M)$ and independent $\forall j \in \{1,\ldots, \Tp\}$.

To estimate $\bf{h}_k$, a channel observation $\bf{z}_k$ is obtained by projecting the received pilot signal $\bf{Y}_\textsc{p}$ onto the pilot used by $\rm{UE}_k$, also known as a de-spreading operation, such that
\begin{equation}
	\bf{z}_k = \bf{Y}_p \frac{\bfs{\phi}_k^*}{\Tp \sqrt{\qul_k}} = \bf{h}_k  + \frac{1}{ \sqrt{\Tp \qul_k}}\left(\bar{\bf{n}} + \bf{D}_{\ep} \bf{D}_h^{\frac{1}{2}} \bar{\bf{r}}\right)
	\label{eq:Channel_obs}
\end{equation}
where ${\bar{\bf{n}} = \bf{N}\bfs{\phi}_k^*/\sqrt{\Tp}\sim \cal{CN}(\bf{0},\sigma^2\bf{I}_M)}$ and ${\bar{\bf{r}} = \sum_{j=1}^{\Tp}\bf{r}_j  [\bfs{\phi}_k]_j^*/\sqrt{\Tp}\sim \cal{CN}(\bf{0},\bf{I}_M)}$ since the Gaussian distribution is invariant to unitary transformations. Note that the distortion term $\bf{D}_h^{\frac{1}{2}} \bar{\bf{r}}$ in \eqref{eq:Channel_obs} depends on the channels and is not Gaussian distributed. Thus, computing the minimum mean squared error (MMSE) channel estimate is very cumbersome. To estimate the channel $\bf{h}_k$, the linear MMSE (LMMSE) channel estimation is used instead, given as follows.
\begin{lemma}
	\label{lem:Ch_est}
	Given the channel observation $\bf{z}_k$ in \eqref{eq:Channel_obs}, the LMMSE channel estimate of $\bf{h}_k$ is
	\begin{equation}
	\label{eq:LMMSE_Ch}
		\hat{\bf{h}}_k =\bb{E}\left\{\bf{h}_k\bf{z}_k^H\right\}\left(\bb{E}\left\{\bf{z}_k\bf{z}_k^H\right\}\right)^{-1} \bf{z}_k= \bf{R}_k \bfs{\Psi}_k^{-1}  \bf{z}_k
	\end{equation}
	with
	\begin{equation}
	\bfs{\Psi}_k = \bf{R}_k + \underbrace{\frac{1}{\Tp \qul_k} \left(\sigma^2 \bf{I}_M + \sum_{i=1}^{K} \qul_i\bf{D}_{\ep}\bf{D}_{\bf{R}_i}\bf{D}_{\ep}\right)}_{\textit{Noise and pilot distortion}}
	\label{eq:ChEst_Psi}
	\end{equation}
	where $\bf{D}_{\bf{R}_i} = \rm{diag}(\bf{R}_i)$. The channel estimation error is $\tilde{\bf{h}}_k = \bf{h}_k - \hat{\bf{h}}_k$, with correlation matrix
	\begin{equation}
	\tilde{\bf{C}}_k = \bb{E}\left\{ \tilde{\bf{h}}_k \tilde{\bf{h}}_k^H \right\} = \bf{R}_k - \bf{R}_k\bfs{\Psi}_k^{-1}\bf{R}_k.
	\end{equation}	
\end{lemma}
\begin{IEEEproof}
	It follows from applying standard LMMSE techniques  \cite[Sec.~3]{Bjornson2017_MaMIMObook}.
\end{IEEEproof}

Due to the HWIs effect in the received pilot signal, the channel estimate $\hat{\bf{h}}_k$ is corrupted by a distortion that depends on the channel of interest $\bf{h}_k$, which cannot be removed by the de-spreading operation in \eqref{eq:Channel_obs}. In addition, the orthogonality between the pilots is destroyed by the HWI effect giving rise to a distortion that depends on the channel realizations from other UEs in the system. To minimize the MSE of the channel estimates, the noise and hardware distortion effect should be as low as possible (corresponding to the last two terms in \eqref{eq:ChEst_Psi}). The term ``pilot distortion'' is precisely used to specify the effect of hardware distortion on the channel estimation.
\begin{remark}
	\label{rem:CHestnotGauss}
	The channel estimates and their errors in Lemma~\ref{lem:Ch_est} are not Gaussian distributed. Thus, they are not statistically independent even though they are uncorrelated by design. This observation is important when deriving the achievable SE expressions later in the paper.
\end{remark}

\subsection{HWI allocation for minimal pilot distortion}
\label{subsec:HWI_pilot_dist}
The main purpose of this article is to study the optimal allocation of hardware quality per antenna to achieve maximal SE or EE. In Massive MIMO, the channel estimates are used to coherently combine the signals from all BS antennas and perform spatial multiplexing. As a result, the quality of channel estimation is crucial to achieve high SE. Thus, a first approach to optimize the allocation of HWI per antenna (i.e, $\ep_m$) is to minimize the pilot distortion. Consider the following optimization problem: 
\begin{align*}
\begin{aligned}
& \minimize_{\ep_m \geq 0,	\; \forall m \in \{1,\ldots,M\}} & & \rm{tr}\left(	\bf{D}_{\ep}    \sum_{i=1}^{K} \qul_i\bf{D}_{\bf{R}_i} \bf{D}_{\ep}\right) \\
&\st_{\hphantom{\ep_m \; \forall m \in \{1,\ldots,M\}}} & & \sum_{m=1}^{M} \log_2\left(\frac{\zeta_m}{\ep_m} \right)\leq b_\textsc{tot}.
\end{aligned}
\stepcounter{equation}\label{eq:OP_MSEChEst}\tag{\theequation}
\end{align*}
The objective minimizes the pilot distortion (see the third term in \eqref{eq:ChEst_Psi}) while the constraint imposes a lower limit on the combined level of HWIs for all antennas. The function that aggregates the level of HWIs across the antennas in the constraint of \eqref{eq:OP_MSEChEst} corresponds to an ADC bit budget $\sum_{m=1}^{M}b_m \leq b_\textsc{tot}\in \bb{N}$, where the ADC bits $b_m$ have been relaxed to be real-valued (see \eqref{eq:ADC_Ce} for mapping an ADC bit resolution $b_m$ to $\ep_m$). 
To make the problem in \eqref{eq:OP_MSEChEst} tractable and find a closed-form solution, there is no constraint to enforce $b_m\geq 1$ which results in positive real-valued ADC bits that can be below 1 (See Figure~\ref{fig:CDF_OpQbitpilotK10}). However, in Section~\ref{sec:int_ADC}, a heuristic method to map the optimal real-valued ADC bit solution to integer values is proposed.

\begin{lemma}
	\label{lem:Convex_op_minPi}
	The problem in \eqref{eq:OP_MSEChEst} has a quadratic objective and a convex constraint since $\log_2(1/x)$ is a convex function. Thus, it is a convex optimization problem of standard form that can be solved efficiently by well established methods \cite{Boyd_cvx_book}. 
\end{lemma}
\begin{remark}
	It is worth mentioning that the ADC bit budget constraint can be changed to another linear or convex constraint, to accommodate other aspects of hardware design, without changing the methodology to solve the problem. As an example, the alternative constraint $\sum_{m=1}^{M} \ep_m \geq \ep_\textsc{tot}$ imposes a total minimum HWI energy $\ep_\textsc{tot}$ that needs to be dispensed. 
\end{remark}
\begin{remark}
Regarding the analysis on the ADC bit allocation, we point out that there is no maximum ADC resolution constraint per BS antenna in \eqref{eq:OP_MSEChEst}. Having an aggregated constraint of an ADC bit budget or total PwC (as shown in Section~\ref{sec:PwC}) would intrinsically limit the maximum ADC bit resolution per BS antenna. Throughout all the simulation results in this paper, the maximum ADC bit resolution that was assigned never surpassed 15 bit, which is common in practical ADCs. Hence, even if we added a maximum ADC bit resolution constraint of 15 bit, it would not impact the results.	
\end{remark}

The optimal HWI allocation is found by analyzing the Lagrange dual problem and evaluating the Karush-Kuhn-Tucker (KKT) conditions. The following theorem summarizes the optimal solution for the optimization problem \eqref{eq:OP_MSEChEst}.

\begin{theorem}
	\label{th:Pilot_dist}
	The optimal solution of the convex optimization problem in \eqref{eq:OP_MSEChEst} is given by
	\begin{align*}
	&\ep_m^\textsc{op} = \left(2^{-b_\textsc{tot}} \prod_{m'=1}^{M}\zeta_{m'}\sqrt{\frac{p_{m'}^\textsc{u}}{p_m^\textsc{u}}} \right)^{\frac{1}{M}}
	 \\
	&\Rightarrow
	b_m^\textsc{op} =\frac{1}{M}\left(b_\textsc{tot} + \sum_{\substack{m'=1\\m'\neq m}}^{M}\log_2\left(\frac{\zeta_m}{\zeta_{m'}} \sqrt{\frac{p_m^\textsc{u}}{p_{m'}^\textsc{u}}}\right)\right)
	\stepcounter{equation}\tag{\theequation} \label{eq:op_ep_pilot_dist}
	\end{align*}
	where $	p_m^\textsc{u} = \sum_{i=1}^{K}\qul_i[\bf{R}_i]_{mm}$ is the average undistorted received pilot power.

\end{theorem}
\begin{IEEEproof} 
	It follows from minimizing the Lagrangian by evaluating the KKT conditions. Since the primal problem in \eqref{eq:OP_MSEChEst} is convex, the primal and dual points that satisfy the KKT conditions are primal and dual optimal with zero duality gap \cite[Ch.~5]{Boyd_cvx_book}. See Appendix~A
	for details.
\end{IEEEproof}
The optimal solution $\ep_m^\textsc{op}$ satisfies the total HWI budget constraint with equality (see Appendix~\ref{app:proof_min_pilot_dist}) which means that the full ADC bit budget $b_\textsc{tot}$ is used. The optimal level of HWIs is allocated based on the received signal strength per antenna, so that higher received power at a given antenna results in lower level of HWI (more ADC bits) allocation and vice-versa. The proportion of HWI allocation is based on the product of the received power across the antennas due to the exponential relationship between the level of HWI $\ep_m$ and the ADC bit $b_m$.  
\begin{remark}
	\label{rem:min_pilot_dist}
	If the average undistorted received pilot power is equal across the antennas (i.e., $p_m^\textsc{u}=p^\textsc{u}$) and all ADCs are equal (i.e., $\zeta_m = \zeta$), \eqref{eq:op_ep_pilot_dist} reduces to equal ADC bit allocation: $b_m^\textsc{op} = \frac{b_\textsc{tot}}{M}$. If these conditions hold approximately, equal ADC resolution is approximately optimal. Similar conclusions were also drawn in \cite{17_Verenzuela_mixed_ADC}.
\end{remark}

\subsection{Integer ADC bit allocation}
\label{sec:int_ADC}
The result in Theorem~\ref{th:Pilot_dist} finds the optimal level of HWI to minimize the pilot distortion. However, when mapping these values into ADC bits $b_m^\textsc{op}$ as shown in \eqref{eq:op_ep_pilot_dist}, these numbers are not likely to be integers. Thus, the heuristic Algorithm~\ref{alg:intalloc} is proposed to map the optimal real-valued ADC bits  $b_m^\textsc{op}$ into integer values that can be implemented in real ADCs. The operator $\lfloor \cdot\rceil$ stands for rounding to the nearest integer. Note that this algorithm allocates the full ADC bit budget $b_\textsc{tot}\in\bb{N}$.

\begin{algorithm}[t]	
	\caption{Procedure to map real-valued ADC bit allocation to integer values.}
	\label{alg:intalloc}
	
		\begin{algorithmic}[1]
			\State Initialize the integer ADC bits as $b_m^\textsc{int} = \lfloor b_m^\textsc{op}\rceil$.
			\State Set  $N^\textsc{diff} = b_\textsc{tot} - \sum_{m=1}^{M} b_m^\textsc{int}$.
			\While {$N^\textsc{diff} \neq 0$}
			\If { $b_m^\textsc{int}<1$ } 
			\State Set to $b_m^\textsc{int}=1$ $\forall m\in \{1,\ldots,M\}$.
			\EndIf
			\State Set  $N^\textsc{diff} = b_\textsc{tot} - \sum_{m=1}^{M} b_m^\textsc{int}$.
			\If {$N^\textsc{diff} > 0$}
			\State Add 1 bit to the $N^\textsc{diff}$ lowest values of $b_m^\textsc{int}$.
			\ElsIf{$N^\textsc{diff} < 0$}
			\State Subtract 1 bit to the $N^\textsc{diff}$ highest values of $b_m^\textsc{int}$.
			\EndIf	
			\EndWhile	 
		\end{algorithmic}
		
\end{algorithm}

\subsection{Numerical example}
\label{subsec:HWIsim_pilot_dist}
To illustrate the impact of spatial channel correlation and LSF variations on the optimal HWI allocation, the model ${\bf{R}_k = \bf{D}_{{\beta}_k}^{\frac{1}{2}}\bar{\bf{R}}_k\bf{D}_{{\beta}_k}^{\frac{1}{2}}}$ is taken as an example for the spatial channel correlation matrices, which is based on \cite{18_Bjornson_unlimited_cap}. The LSF coefficients between $\rm{UE}_k$ and each BS antenna element are enclosed in $\bfs{\beta}_k = [\beta_{1k},\ldots,\beta_{Mk}]^T$, such that $\bf{D}_{\bfs{\beta}_k} = \rm{diag}(\bfs{\beta}_k)$. The correlation matrix $\bar{\bf{R}}_k$ has ${[\bar{\bf{R}}_k]_{mm} = 1}$ and $|[\bar{\bf{R}}_k]_{mm'}|\leq 1 \;\forall m\neq m' \in \{1,\ldots,M\}$ to account for the correlation between antenna elements.  Recall that $\bar{\beta}_k = \rm{tr}(\bf{R}_k)/M$  is the average channel gain. Four cases are simulated for the spatial correlation matrices:

	\begin{enumerate}
		\item \textit{Co-located antennas and spatial correlation (Co-corr-I)}: The matrix $\bar{\bf{R}}_k$ follows the Gaussian local scattering model  defined in \cite[Ch.~2]{Bjornson2017_MaMIMObook}, with angular standard deviation $\sigma_\rm{ang}$. The LSF coefficients are equal for all antenna elements such that ${\rm{diag}(\bfs{\beta}_k) = \bar{\beta}_k\bf{I}_M}$. The average channel gain is modeled as $\bar{\beta}_k = \omega^{-1} d_{k}^{-\alpha} \digamma^\rm{sh}$, where $\omega$ is the fixed pathloss at 1~km accounting for fixed propagation effects (e.g., wall penetration), $\alpha$ is the pathloss exponent, and $d_{k}$ is the distance between $\rm{UE}_k$ and the BS (measured in [km]). The shadow fading effects are included in $\digamma^\rm{sh}$ and modeled as independent log-normal distributed with standard deviation $\sigma_\rm{sh}$ [dB], that is, $10 \log_{10}(\digamma^\rm{sh})\sim \cal{N}(0,\sigma_\rm{sh}^2)$. Note that since the antennas are co-located the distance $d_{k}$ is considered to be the same for all antennas at the BS.
		
		\item \textit{Co-located antennas, spatial correlation and rank 1 diagonal variations (Co-corr-D1)}: The matrix $\bar{\bf{R}}_k$ is defined as in the previous case, whilst the variations of the LSF coefficients of each BS antenna are included following a log-normal distribution with standard deviation $\sigma_\rm{lsf}$. Thus, $\beta_{mk} = \bar{\beta}_k \digamma_m^\rm{lsf}$ where $10 \log_{10}(\digamma_m^\rm{lsf})\sim \cal{N}(0,\sigma_\rm{lsf}^2)$. Note that the term ``rank 1'' means that the same realization of $\digamma_m^\rm{lsf}$ is shared for all $K$ UEs.
		 	
		\item \textit{Co-located antennas, spatial correlation and rank $K$ diagonal variations (Co-corr-DK)}: The matrix $\bar{\bf{R}}_k$ is defined as in the case of \textit{Co-corr-I}, and the variations across the LSF coefficients of each BS antenna are modeled following a log-normal distribution with standard deviation $\sigma_\rm{lsf}$. Thus, $\beta_{mk} = \bar{\beta}_k \digamma_{mk}^\rm{lsf}$ where $10 \log_{10}(\digamma_{mk}^\rm{lsf})\sim \cal{N}(0,\sigma_\rm{lsf}^2)$. The term ``rank $K$'' means that each UE and antenna element have variations in the LSF with independent realizations.
		
		\item \textit{Cell-free and uncorrelated channels (Cell-free)}: The matrix $\bar{\bf{R}}_k = \bf{I}_M$ and the antenna elements are assumed to be located at different positions so that $\beta_{mk} = \omega^{-1} d_{mk}^{-\alpha} \digamma^\rm{sh}$, where $d_{mk}$ corresponds to the distance between $\rm{UE}_k$ and the $m^{th}$ antenna element. The shadow fading effects are included in $\digamma^\rm{sh}$ and are modeled as in the case \textit{Co-corr-I}. 
	\end{enumerate}

The UEs are dropped uniformly at random in a squared area. In the case of co-located antennas, the BS is located at the center of the square whereas in the \textit{Cell-free} case the antennas are distributed uniformly at random. A minimum distance of 10~meters is enforced between all UEs and BS antennas. Statistical channel inversion power control is assumed such that ${\qul_k = \min \left\{\rho_\textsc{max},\,\bar{\qul}/  \bar{\beta}_k\right\}}$ where $\bar{\qul}$ is used to set the SNR level and $\rho_\textsc{max}$ is the maximum transmission energy per symbol.
Table~\ref{tab:sim_param} summarizes the main simulation parameters used.

\begin{table}[!t]		
	\centering
	\captionsetup{width=0.45\textwidth}
	\caption{Simulation parameters.}
	\label{tab:sim_param}	
	{
		\begin{tabular}{@{\hskip -2pt}C{4.32cm}@{\hskip 0pt}|@{\hskip 0pt}C{4.6cm}@{\hskip -2pt}}
			\bfseries Parameter  & \bfseries Value\\
			\hline
			System bandwidth & $\Bw = 20$ [MHz] \\ 

			Max. trans. power per UE & $10\log_{10}(\rho_\textsc{max}\Bw) = 20$ [dBm]\\

			Noise power & $10 \log_{10}\left(\sigma^2\Bw\right) = -94$ [dBm] \\ 			

			Square side length & $0.4$ [km] \\	
 			
			Pathloss exponent & $\alpha = 3.76$  \\
 		
			Pathloss at $1$ km & $\omega = 148.1$ [dB]  \\ 			

			Shadow fading std. dev.& $\sigma_{\rm{sh}}=10$ [dB]\\ 
 								 		
			Angular std. deviation & $\sigma_\rm{ang} = 10^\circ$\\
 								 		
			Antenna variations std. dev.& $\sigma_{\rm{lsf}}=4$ [dB]\\ 
 								 		
			ADC constant& $\zeta_m=1.6$ $\forall m\in\{1,\ldots,M\}$\\ 

			Power amplifier efficiency & $\eta = 0.39$  \\ 
			Fixed circuit power & $\rm{P}_\textsc{cst} = 10$ [W] \\ 			

			coding/decoding power  & $\rm{P}_\textsc{cd} = 1.15$ [Joule/Gbit] \\ 
			Power per UE  & $\rm{P}_\textsc{ue} = 0.1$ [W] \\ 
			Power per BS antenna  & $\rm{P}_\textsc{bs-a} = 0.05$ [W] \\														
			\hline 			
		\end{tabular}		}	
	\end{table}

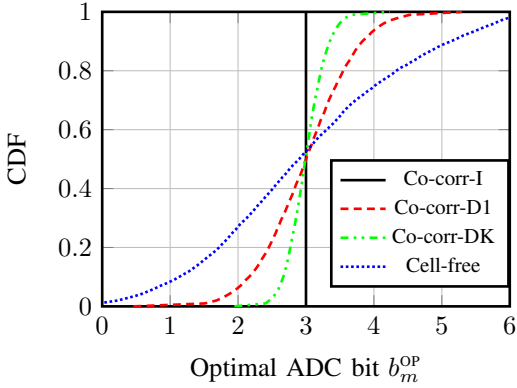
\begin{figure}[!t]		
	\centering
	\captionsetup{width=0.45\textwidth}
	\import{Figures/}{CDF_OpQbit_PilotDistK10.tex}
	\caption{\small CDF of optimal ADC bit allocation with $M = 100$, $K = 10$, $\bar{\qul}/\sigma^2 = 1$ \mbox{(SNR $= 0$ [dB])}, and $b_\textsc{tot} = 3 M$.}
	\label{fig:CDF_OpQbitpilotK10}	
\end{figure}

Figure~\ref{fig:CDF_OpQbitpilotK10}  shows the empirical cumulative distribution function (CDF) of the optimal allocation of ADC bits given in Theorem~\ref{th:Pilot_dist}. In the \textit{Co-corr-I} case there are no LSF variations across antenna elements which yields equal ADC bits as the optimal allocation. In contrast, in the other cases, as more variations are introduced in the diagonal elements of the spatial correlation matrices, the optimal ADC bit allocation becomes more spread.

	\section{Achievable SE}
	\label{sec:AchSE}
To send the information through the channel, the transmitted signal from $\rm{UE}_k$, in \eqref{eq:rx_signal}, is defined as ${x_k = \sqrt{\pul_k}s_k}$ where the unit-power data symbol is $s_k\sim \cal{CN}(0,1)$ and $\pul_k$ is the average energy per data symbol. At the BS, linear signal processing is assumed to combine the signals from all BS antennas, such that, $\hat{s}_k=\bf{v}_k^H\bf{y}$ is the observation of the data symbol $s_k$ given by
\begin{align*}
\hat{s}_k 
  =& \underbrace{\vphantom{\sum_{\substack{i\neq k}}^{K}}
		  	     \bb{E}\!\left\{\bf{v}_k^H\bf{h}_k\right\}\! \sqrt{\pul_k}  s_k}_{\textit{Desired signal}} 
  \!+\! \underbrace{\vphantom{\sum_{\substack{i\neq k}}^{K}} 
				 \left(\bf{v}_k^H\bf{h}_k \!-\! \bb{E}\!\left\{\bf{v}_k^H\bf{h}_k\right\}\!\right)\!\sqrt{\pul_k} s_k}_{\textit{Channel gain uncertainty}}
  \\
  &
  \!+ \!\underbrace{\sum_{\substack{i\neq k}}^{K}\!\bf{v}_k^H\bf{h}_i\sqrt{\pul_i} s_i}_{\textit{Inter-user interference}} 
  \!+\! \underbrace{\vphantom{\sum_{\substack{i\neq k}}^{K}}
  	            \bf{v}_k^H\bf{n}}_{\textit{Noise}} 
  \!+\!\! \underbrace{\vphantom{\sum_{\substack{i\neq k}}^{K}}
			  	\bf{v}_k^H\bf{e}}_{\textit{Data distortion}}
  \stepcounter{equation} \tag{\theequation}\label{eq:data_obs}
\end{align*}
where the term ``data distortion'' in \eqref{eq:data_obs} refers to the negative effect that hardware distortion has on	 the combined data signal. The combining vector is given by $\bf{v}_k$ and two different options are considered:
\begin{align*}
\bf{v}_k =
	\begin{cases}
	\hat{\bf{h}}_k  & \text{MR}\\[-5pt]
	\left( \sum_{i = 1}^{K} \hat{\bf{h}}_i\hat{\bf{h}}_i^H\pul_i + \sigma^2\bf{I}_M \right)^{-1} \hat{\bf{h}}_k\pul_k  & \text{RZF}.
	\end{cases}
\end{align*}  
MR combining requires the least amount of operations to implement, and it maximizes the gain of the desired signal. Regularized zero-forcing (RZF) is the state-of-the-art heuristic scheme for this setup. It
adds a level of complexity by introducing a matrix inversion but suppresses inter-user interference, while maintaining a decent SNR level for the desired signal.

The achievable SE is calculated as a lower bound on the ergodic capacity, which is obtained by using the use-and-then-forget methodology assuming Gaussian data symbols and worst case independent Gaussian effective noise \cite{Bjornson2017_MaMIMObook, 16_Marzetta_MAMIMO_book}. The effective noise is composed of the channel gain uncertainty, inter-user interference, noise and data distortion effects as shown in \eqref{eq:data_obs}. Note that the channel gain uncertainty term is a consequence of the capacity lower bound and for large number of BS antennas, as in the case of Massive MIMO, it has a negligible impact on the tightness of the lower bound \cite{Bjornson2017_MaMIMObook, 16_Marzetta_MAMIMO_book}. The achievable SE is given in the following theorem. 
\begin{theorem}
	\label{th:AchSE}
	An achievable SE for $\rm{UE}_k$ $\forall k \in \{1,\ldots,K\}$ under hardware distortions is 
\begin{equation}
\rm{SE}_k = \left(1 - \frac{\Tp}{\Tc}\right) \log_2\left(\vphantom{\frac{K^K}{K^K}} \right. 1 + \underbrace{\frac{ \left|\bb{E}\left\{  \hat{s}_k s_k^*\right\}\right|^2}{\bb{V}\sf{ar} \left(\hat{s}_k  -  \bb{E}\left\{  \hat{s}_k s_k^*\right\}s_k\right)  }}_{=\rm{SINR}_k}\left.\vphantom{\frac{K^K}{K^K}} \right)
\label{eq:Ach_SE}
\end{equation}
where $\bb{V}\sf{ar}(\cdot)$ is the variance operator. The effective SINR with the HWI model in \eqref{eq:distortion_e} is
\begin{align*}
\label{eq:SINR_all_v}
\rm{SINR}_k&= {\left|\bb{E}\left\{\bf{v}_k^H\bf{h}_k\right\}   \right|^2 \pul_k} \Bigg / \Bigg ( 
\sum\limits_{i=1}^{K}\bb{E}\left\{\left|\bf{v}_k^H\bf{h}_i\right|^2\right\}\pul_i
\\
& - \left|\bb{E}\left\{\bf{v}_k^H\bf{h}_k\right\}   \right|^2 \pul_k
+ \bb{E}\left\{\left|\bf{v}_k^H\bf{n}\right|^2 \right\}+ \bb{E}\left\{\left|\bf{v}_k^H\bf{e}\right|^2 \right\}
\Bigg ).
\stepcounter{equation}\tag{\theequation}
\end{align*}
\end{theorem}
\begin{IEEEproof}
	The achievable SE in \eqref{eq:Ach_SE} follows from applying the use-and-then-forget bound \cite{Bjornson2017_MaMIMObook, 16_Marzetta_MAMIMO_book} on the equivalent SISO channel in \eqref{eq:data_obs} with deterministic gain $\bb{E}\{\bf{v}_k^H\bf{h}_k\}\sqrt{\pul_k}$ and non-Gaussian noise.
\end{IEEEproof}
The achievable SE in \eqref{eq:Ach_SE} can be used for any type of hardware distortion, in particular, exact quantization which is used later on to validate the model in \eqref{eq:distortion_e}. The effective SINR \eqref{eq:SINR_all_v} follows by assuming the model in \eqref{eq:distortion_e} and it is valid for all linear combining methods. In the case of MR processing, the effective SINR can be found in closed-form as shown in the following corollary.
\begin{corollary}
	\label{cor:MR_SE_CL}
In the case of MR combining, a closed-form expression for the effective SINR is given in \eqref{eq:SINR_MR_mat} at the top of next page.
\begin{figure*}[t]
\begin{align*}
\overline{\rm{SINR}}_k =& \pul_k\rm{tr}\left(\bf{R}_k\bfs{\Psi}_k^{-1}  \bf{R}_k\right)^2  \Bigg/ \left(\vphantom{\sum_K^K}\right.
\underbrace{ \vphantom{\sum}
\pul_k\rm{tr}\left(\bf{D}_{\ep}^2\rm{abs}\left(\rm{diag}\left(\bf{R}_k\bfs{\Psi}_k^{-1}  \bf{R}_k \right)\right)^2 \right)
}_{\textit{Self-distortion ($\rm{UE}_k$)}}
\\
&\mkern 0mu
+	\sum_{i=1}^{K}\frac{\pul_i \qul_i}{\Tp \qul_k}
	 \rm{tr}\left( \vphantom{\sum\nolimits_k} \right.
\underbrace{\vphantom{\sum\nolimits_k^k}
	 \bf{D}_{\ep}^2\rm{abs}\left(\rm{diag}\left(\bf{R}_i\bf{R}_k\bfs{\Psi}_k^{-1} \right)\right)^2 
}_{\textit{Distortion from inter-user interference}}
	+
 \underbrace{\vphantom{\sum\nolimits_k^k}
	  \bf{D}_{\ep}^2  \rm{diag}\left(\rm{abs}\left(\bf{R}_k\bfs{\Psi}_k^{-1}\right)^2 \bf{D}_{\ep}^2  \rm{abs}\left(\bf{R}_i\right)^2 \right)
}_{\textit{Additional distortion from all UEs}}
      \left.\vphantom{\sum\nolimits_k} \right)
\\
&\mkern 0mu
+  \rm{tr}\left( \vphantom{\sum\nolimits_{k}^{k}}\right. 
\left( \vphantom{\sum\nolimits_{k}^{k}}\right.
\underbrace{\vphantom{\sum\limits_{k}^{k}}
	\sum_{i=1}^{K}\pul_i\bf{R}_i}_{\textit{Inter-user interference}}  
+ \underbrace{\vphantom{\sum\limits_{k}^{k}}
	\sigma^2\bf{I}_M}_{\textit{Noise}} 
+ \underbrace{\vphantom{\sum\limits_{k}^{k}}
	\sum_{i = 1}^{K}\pul_i \bf{D}_{\ep}^2 \bf{D}_{\bf{R}_i}}_{\textit{Data distortion}} 
\left. \vphantom{\sum\nolimits_{k}^{k}}\right)
\bf{R}_k\bfs{\Psi}_k^{-1}  \bf{R}_k 
\left. \vphantom{\sum\nolimits_{k}^{k}}\right)
\left. \vphantom{\sum\limits_{K}^{K}}\right)
\stepcounter{equation}\tag{\theequation}\label{eq:SINR_MR_mat}
\end{align*}
\hrulefill
\end{figure*}
\end{corollary}
\begin{IEEEproof}
	The proof is provided in  Appendix~B.
\end{IEEEproof}

The effect of HWIs on the SE is more complicated than in the channel estimates as it is shown in the SINR expression in \eqref{eq:SINR_MR_mat}. Since the hardware distortion affects both the channel estimates and data signals, cross products arise in the linear combining process which, in turn, form the additional distortion terms shown in \eqref{eq:SINR_MR_mat}. To illustrate how the spatial correlation and large-scale variations along the antennas affect the SE, numerical results are presented next.

\subsection{Numerical example with minimum pilot distortion}
\label{subsec:sim_pilot_dist}

To validate the accuracy of the results obtained with the HWI model in \eqref{eq:distortion_e}, the SE with exact quantization is also calculated, such that the quantized received signal for an arbitrary sample of the coherence block is  $\bf{y}_q = \bb{Q}(\sum_{i = 1}^{K}  \bf{h}_i x_i + \bf{n})$, where $\bb{Q}(\cdot)$ stands for the quantization operation. For ADC bit resolutions of up to five $b_m\leq 5$, the quantization levels are optimized as in \cite{J_Max_Q_min_dist}, and for  $b_m> 5$ uniform quantization is used with optimized levels as in \cite{D_Hui_D_Neuhoff_asymQ}. The channel estimates are obtained following the same procedure as in Section~\ref{sec:Ch_est}, where the quantized received pilot signal corresponds to the $\tau_p$ instances of $\bf{y}_q$ with pilot symbols. The LMMSE channel estimates are given by the first equality in \eqref{eq:LMMSE_Ch}, where the expectations are found numerically. The SE with exact quantization is computed using \eqref{eq:Ach_SE}, where the data estimates are given by the first equality in \eqref{eq:data_obs} considering quantized channel estimates and received data signal. In all the results with exact quantization, the integer bit allocation algorithm proposed in Section~\ref{sec:int_ADC} is used.

\begin{figure*}[!t]
	{
		\captionsetup{width=0.45\textwidth}
		\centering
		\subfloat[MR and equal ADC bit allocation.]{\import{Figures/}{CDF_EqQbit_MR_2020.tex}		%
			\label{fig:CDF_EqQbit_MR}}
		\subfloat[RZF and equal ADC bit allocation.]{\import{Figures/}{CDF_EqQbit_RZF_2020.tex}		%
			\label{fig:CDF_EqQbitRZF}}
		\\[-10pt]
		\subfloat[MR and min pilot distortion ADC bit allocation.]{\import{Figures/}{CDF_minPilotQbit_MR_2020.tex}		%
			\label{fig:CDF_minPilotQbitMR}}
		\subfloat[RZF and min pilot distortion ADC bit allocation.]{\import{Figures/}{CDF_minPilotQbit_RZF_2020.tex}		%
			\label{fig:CDF_minPilotQbitRZF}}
	}
	\caption{CDF of SE per UE for $M=100$, $K = 10$, $\Tc = 200$, $\bar{\qul}/\sigma^2 = 1$ \mbox{(SNR $= 0$ [dB])}, and $b_\textsc{tot} = 3 M$. The lines correspond to the SE with the additive distortion model in \eqref{eq:distortion_e}, whereas the markers are the results with exact quantization and integer bit allocation (see Section~\ref{sec:int_ADC}).}
	\label{fig:CDF_SE_Eq_minPilot_Qbit}
\hrulefill
\end{figure*}
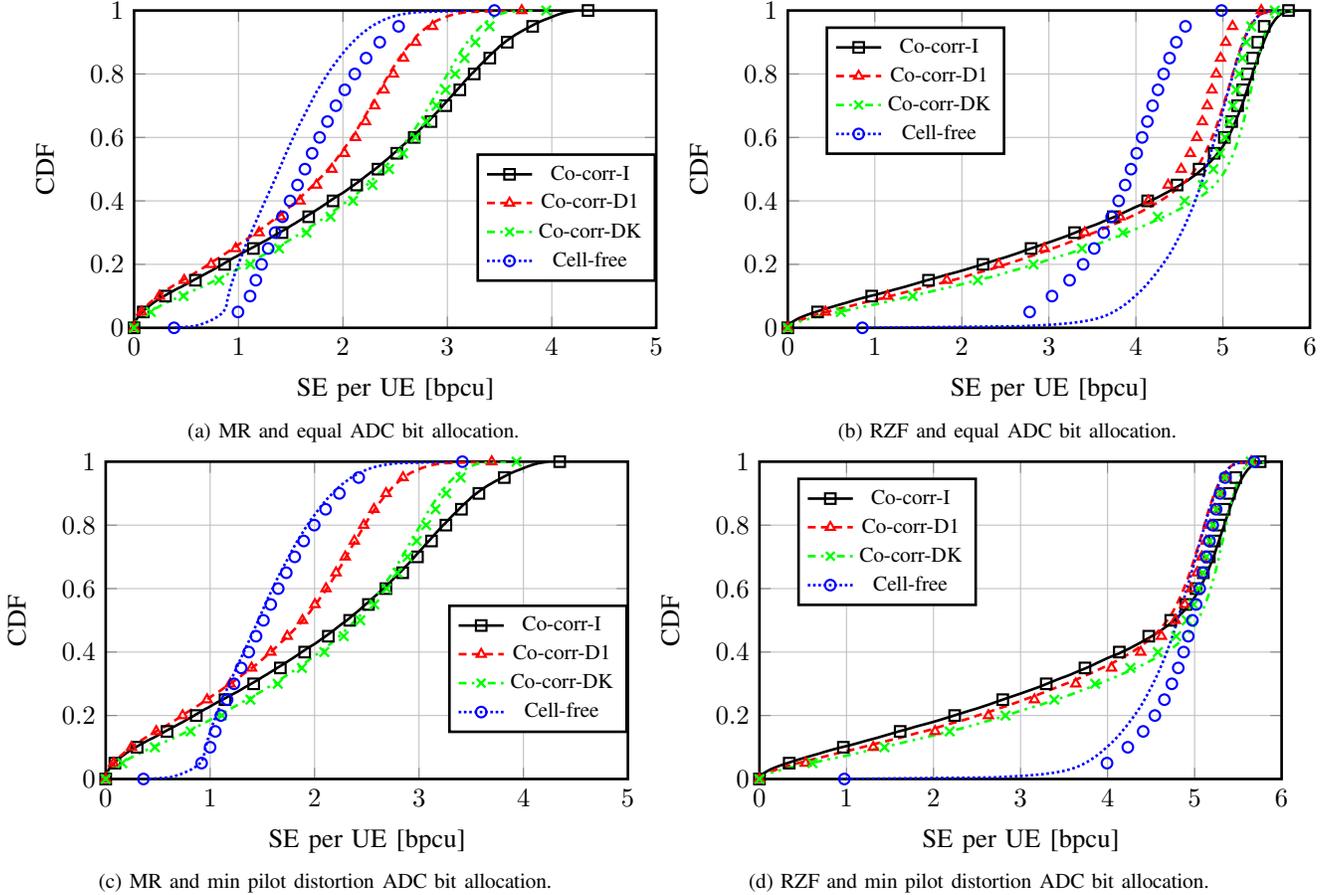

Figure~\ref{fig:CDF_SE_Eq_minPilot_Qbit} shows empirical CDF curves of the SE per UE where the spatial correlation matrices are modeled based on the four cases in Section~\ref{subsec:HWIsim_pilot_dist}. The markers correspond to results with the exact quantization model and the lines correspond to \eqref{eq:SINR_all_v} from Theorem~\ref{th:AchSE} for RZF, and \eqref{eq:SINR_MR_mat} from Corollary~\ref{cor:MR_SE_CL} for MR. It can be seen that the additive distortion model in \eqref{eq:distortion_e} is accurate in most cases. However, in the \textit{Cell-free} case with equal ADC bit allocation, there is a gap between the exact quantization and the additive distortion model in \eqref{eq:distortion_e}. 
A possible explanation is that the distortion becomes practically independent when it is combined from many different signals that have similar strength, which might not be the case in \textit{Cell-free} since the UEs are located at different distances to the antennas. This discrepancy is substantially reduced when the ADC bit allocation is optimized to minimize the pilot distortion since the antennas with higher received power are allocated more ADC bits that allows them to increase their dynamic range. 

In the case of MR combining, optimizing the ADC bit allocation provides almost no SE gains. In contrast, when using RZF, improvements of up to 1 [bit-per-channel-use] ([bpcu]) per UE for the \textit{Cell-free} case with the exact quantization model are shown in Figure~\ref{fig:CDF_minPilotQbitRZF}. This indicates that the optimal ADC bit allocation is mostly beneficial for obtaining accurate channel estimates that enhance the performance of interference suppression techniques like RZF.

\section{Optimal HWI allocation}
\label{sec:Op_HWI_allc}

In Section~\ref{subsec:HWI_pilot_dist}, the optimal HWI allocation for minimal pilot distortion was found in closed form. This result gave good insights into how to allocate the HWI based on the received signal strength of the pilot signals. However, minimizing the pilot distortion does not necessarily maximize the performance in terms of SE since the impact of HWIs is fundamentally different as it is shown in Theorem~\ref{th:AchSE}. The same is true for the PwC where less distortion often comes at a higher PwC (e.g., a higher ADC bit resolution increases the PwC). To further study the optimal allocation of the HWI level per antenna, consider the following optimization framework:
\begin{align*}
&\maximize_{\substack{\bfs{\ep} = [\ep_1,\ldots,\ep_M]^T \\\bfs{\pul} = [\pul_1,\ldots,\pul_K]^T}} 			&  	&\mkern 0mu f^\textsc{obj}(\bfs{\ep},\bfs{\pul})\\
& \st_{\hphantom{\substack{\bfs{\pul} = [\pul_1,\ldots,\pul_K]^T}}}	&	&\mkern 0mu f^\textsc{hwi}(\bfs{\ep},\bfs{\pul})\leq 0
,
\\ & & & 
0\leq \ep_m \leq \ep_{\textsc{max}} \: \forall m \in \{1,\ldots,M\},
\\
& &	&\mkern 0mu
0\leq \pul_k \leq \pul_{\textsc{max}} \: \forall k \in \{1,\ldots,K\},
\stepcounter{equation}\tag{\theequation} \label{eq:general_OP}
\end{align*}
where $\pul_{\textsc{max}}$ is the maximal transmission energy per symbol and $\ep_{\textsc{max}}$ is the maximal HWI level so that $\ep_m \leq \ep_{\textsc{max}}\,\Rightarrow\,b_m\geq 1$. 

The optimization problem formulation in \eqref{eq:general_OP} constitutes a general framework to optimize the level of HWI per antenna and the transmission energy per data symbol. In uplink communications, power control is essential to avoid signals from UEs with low channel gains to be overshadowed by signals from UEs with high channel gains. In addition, power control is important to reduce the dynamic range of signals so that the ADCs introduce less quantization distortion. Thus, it is desirable to optimize the transmission energy per data symbol and the level of HWI together. The function $f^\textsc{obj}(\bfs{\ep},\bfs{\pul})$ corresponds to the optimization objective and it can be set to maximize the SE, or another useful utility function, such as minimizing the distortion or maximizing the EE. The function $f^\textsc{hwi}(\bfs{\ep},\bfs{\pul})$ enforces a constraint on the HWI level and (or) transmission energy per data symbol based on design requirements, such as, an ADC budget, or maximum PwC, among others.

The selection of $f^\textsc{obj}(\bfs{\ep},\bfs{\pul})$ and $f^\textsc{hwi}(\bfs{\ep},\bfs{\pul})$ depends on the design requirements of the system. However, it is also important to choose functions that allow for amenable formulations that can be solved efficiently. In the following sections, the optimization framework in \eqref{eq:general_OP} is used to obtain insights into the optimal HWI allocation to maximize the SINR under an ADC bit budget or a PwC constraint. The functions $f^\textsc{obj}(\bfs{\ep},\bfs{\pul})$ and $f^\textsc{hwi}(\bfs{\ep},\bfs{\pul})$ are selected so that \eqref{eq:general_OP} can be rewritten as geometric programs for which efficient solvers exist \cite{Boyd_cvx_book}.

\subsection{Maximizing the SINR}
\label{subsec:max_SINR}

This section aims at acquiring insights into the optimal HWI allocation that maximizes the SE with MR combining. As shown in Theorem~\ref{th:AchSE}, the SE has a logarithmic dependency on the effective SINR. Since the logarithm is a concave and monotonically increasing function, it is often advantageous to optimize the effective SINRs directly to find more tractable formulations.
To define the objective function $f^\textsc{obj}(\bfs{\ep},\bfs{\pul})$, it is necessary to find a good compromise between fairness and aggregated performance of the UEs. Thus, consider the following objectives
\begin{equation}
f^\textsc{obj}(\bfs{\ep},\bfs{\pul}) = \begin{cases}
  \prod\limits_{\substack{k=1\\[-8pt]\hphantom{k\,\in \{1,\ldots,K\}}   }}^{K} \overline{\rm{SINR}}_{k}(\bfs{\ep},\bfs{\pul}) & \textit{ Max-prod SINRs}\\
  \min\limits_{k\,\in \{1,\ldots,K\}} \overline{\rm{SINR}}_{k}(\bfs{\ep},\bfs{\pul}) & \textit{ Max-min fairness}.
\end{cases}
\label{eq:SINR_obj}
\end{equation}  

\begin{figure*}[t]
	\begin{align*}
			f_k^\textsc{d}(\bf{x},\bfs{\pul})	 =& \sum_{m = 1}^{M}\Bigg( 
	\pul_k\left|\left[\bf{A}_k\right]_{m m}\right|^2
	+ \sum_{i=1}^{K}\left(\frac{\pul_i \qul_i}{\Tp \qul_k} \left|\left[\bf{R}_i\bf{R}_k\bfs{\Psi}_k^{-1} \right]_{mm}\right|^2 + \pul_i \left[\bf{R}_i\right]_{mm} \left[\bf{A}_k \right]_{m m}\right) \Bigg)x_m
	\\
	&
	+ \bf{x}^T\left(
	\sum_{i=1}^{K}\frac{\pul_i \qul_i}{\Tp \qul_k}\rm{abs}\left(\bf{R}_k\bfs{\Psi}_k^{-1}\right)^2 \circ  \rm{abs}\left(\bf{R}_i\right)^2
	\right)\bf{x}
	+\rm{tr}\left(\left(\sum_{i=1}^{K}\pul_i\bf{R}_i + \sigma^2\bf{I}_M\right)\bf{A}_k\right).
	\tag{20}\label{eq:f_k_denSINR}
	\end{align*}
	\hrulefill
\end{figure*}

The first objective in \eqref{eq:SINR_obj} maximizes the product of SINRs (i.e., the geometric mean) and it aims at maximizing the aggregated performance of all UEs, while maintaining a certain level of fairness since UEs with very low SINR would make the objective function small. This objective is a lower bound on the sum SE in which the term ``1'' is removed from the SE expression \eqref{eq:Ach_SE}, (see \cite[Ch.~7]{Bjornson2017_MaMIMObook} for more details). The second objective in \eqref{eq:SINR_obj} is called max-min fairness and it seeks to maximize the lowest SINR among all UEs. Here, the focus is on balancing the SE among the UEs to create the highest even performance as possible. The function
\begin{equation}
f^\textsc{hwi}(\bfs{\ep}) = \sum_{m=1}^{M} \log_2\left(\frac{\zeta_m}{\ep_m} \right) - b_\textsc{tot}
\label{eq:f_ADCbudget}
\end{equation}
is set to satisfy an ADC budget constraint as in Section~\ref{subsec:HWI_pilot_dist}. Notice that in the uplink, each UE has an independent power budget, thus, there is no need to add a constraint to the aggregated transmission energy per data symbols, which means that $f^\textsc{hwi}(\bfs{\ep})$ does not depend on $\bfs{\pul}$.

The effective SINR in \eqref{eq:SINR_MR_mat} depends on $\bfs{\Psi}_k$, which in turn, is a function of the HWI level through the pilot distortion term in \eqref{eq:ChEst_Psi}. As a result, the problem \eqref{eq:general_OP} with the objectives given in \eqref{eq:SINR_obj} and constraint in \eqref{eq:f_ADCbudget}, are not convex. The impact of HWI in $\bfs{\Psi}_k$ is encompassed by the pilot distortion which was analyzed in Section~\ref{subsec:HWI_pilot_dist}. Thus, to gain more insights into the HWI level allocation for maximal SINR, the optimization problems can be solved by assuming $\bfs{\Psi}_k$ is fixed. In this case, the problem for both objectives in \eqref{eq:SINR_obj} and constraint in \eqref{eq:f_ADCbudget} can be casted as geometric programming problems. This result is summarized in the following theorem.

 \begin{theorem}
 	\label{th:Geo_OP_SE}
 	Consider the variable change 	${x_m = \ep_m^2}$ ${\forall m \in \{1,\ldots,M\}}$ and ${\bf{x}= [x_1,\ldots,x_M]^T}$. If $\bfs{\Psi}_k$ is fixed, then 
 	\eqref{eq:general_OP} with the objectives given in \eqref{eq:SINR_obj} and constraint in \eqref{eq:f_ADCbudget}, can be casted as geometric programming problems with the following formulations:

{\underline{Max-prod SINRs}} 	 
 	 \begin{equation}
 	 \begin{aligned}
 	 &\maximize_{\substack{t\geq 0,\:x_m\geq 0,\;\pul_k\geq 0,\\\forall m\in\{1,\ldots,M\}\\\forall k\in\{1,\ldots,K\} }} 			&  	&\mkern -30mu t\prod_{k=1}^{K}\pul_k w_k   	\\[-3pt]
 	 & \st_{\hphantom{\substack{t\geq 0,\:x_m\geq 0,\;\pul_k\geq 0,\\\forall m\in\{1,\ldots,M\}\\\forall k\in\{1,\ldots,K\} }}}	&	&\mkern -30mu t\prod_{k=1}^{K} f_k^\textsc{d} (\bf{x},\bfs{\pul})\leq 1,\\
 	 & 														&	&\mkern -30mu  \prod_{m=1}^{M}\zeta_m^2 x_m^{-1} \leq 2^{2b_\textsc{tot}},\\[-2pt]
 	 &														&	&\mkern -30mu   x_m \leq \ep_{\textsc{max}}^2,\;\pul_k\leq \pul_\textsc{max}.
 	 \end{aligned}
 	 \label{eq:OP_maxprod_fxPsi}
 	 \end{equation}
 	 
{\underline{Max-min fairness}}  	 
 	 \begin{equation}
 	 \begin{aligned}
 	 &\maximize_{\substack{t\geq 0,\:x_m\geq 0,\;\pul_k\geq 0,\\\forall m\in\{1,\ldots,M\}\\\forall k\in\{1,\ldots,K\} }} 			&  	&\mkern -30mu t 	\vphantom{\prod_{k=1}^{K}}\\[-3pt]
 	 & \st_{\hphantom{\substack{t\geq 0,\:x_m\geq 0,\;\pul_k\geq 0,\\\forall m\in\{1,\ldots,M\}\\\forall k\in\{1,\ldots,K\} }}} 	&  	&\mkern -30mu   \frac{tf_k^\textsc{d}(\bf{x},\bfs{\pul})}{ \pul_k w_k  }\leq 1,	\\
 	 &															&	&\mkern -30mu \prod_{m=1}^{M}\zeta_m^2 x_m^{-1} \leq 2^{2b_\textsc{tot}},\\[-2pt]
 	 &															&	&\mkern -30mu  x_m \leq \ep_{\textsc{max}}^2,\;\pul_k\leq \pul_\textsc{max},
 	 \end{aligned}
 	 \label{eq:OP_maxmin-fxPsi}
 	 \end{equation} 	  	 
 	 where $w_k = \rm{tr}\left(\bf{A}_k\right)^2$, $\bf{A}_k = \bf{R}_k\bfs{\Psi}_k^{-1}  \bf{R}_k $ and $	f_k^\textsc{d}(\bf{x},\bfs{\pul})$ is defined in \eqref{eq:f_k_denSINR} at the top of this page.
	\stepcounter{equation}
 \end{theorem}
 \begin{IEEEproof}
It follows from rewriting the effective SINR in \eqref{eq:SINR_MR_mat} as $\overline{\rm{SINR}}_k = \pul_k w_k/f_k^\textsc{d}(\bf{x},\bfs{\pul})$, and noticing that $f_k^\textsc{d}(\bf{x},\bfs{\pul})$ is a posynomial function of $x_m$ and $\pul_k$. Then, the max-product of SINRs problem can be re-casted as in \eqref{eq:OP_maxprod_fxPsi} by adding a new variable $t$  to impose a constraint on the product of the SINR denominators. The max-min fairness problem can be directly rewritten as in \eqref{eq:OP_maxmin-fxPsi} by doing an epigraph formulation \cite{Boyd_cvx_book}. Finally, it can be seen that both problems \eqref{eq:OP_maxprod_fxPsi} and  \eqref{eq:OP_maxmin-fxPsi} maximize a monomial with constraints being posynomials or monomials lower than a constant. Thus, they are geometric programs \cite{Boyd_cvx_book}.
 \end{IEEEproof}
 
To account for the dependency that $\bfs{\Psi}_k$ has on $\ep_m$, the iterative Algorithm~\ref{alg:updatepilotDist} is proposed. Notice that the proposed algorithm converges because the optimization problem always maximizes the same objective. However, since $\bfs{\Psi}_k$ is kept fixed in each optimization step, it is not guaranteed that the solution converges to the global optimum.   

\begin{algorithm}[t]		
	\caption{Iterative procedure to update the pilot distortion $\bfs{\Psi}_k$.}

	\label{alg:updatepilotDist}
 \begin{algorithmic}[1]
 	\State Initialize $\bfs{\Psi}_k$ by setting ${\ep_m = \zeta_m 2^{-\frac{b_\textsc{tot}}{M}}}$ ${\forall m\in \{1,\ldots,M\}}$.
 	\Repeat 
 	\State Solve the optimization problem \eqref{eq:OP_maxprod_fxPsi} or  \eqref{eq:OP_maxmin-fxPsi} and store the optimal solution $\ep_m^\textsc{op-se}$. 
 	\State Update  $\bfs{\Psi}_k$ by setting $\ep_m = \ep_m^\textsc{op-se}$. 
 	\Until{Convergence is achieved.} 
 \end{algorithmic}

 \end{algorithm}

 \begin{figure*}[!t]
 	{
 		\captionsetup{width=0.45\textwidth}
 		\centering
 		\subfloat[{Optimal ADC bit allocation with $\Tc = 200$.}]{\import{Figures/}{CDF_Qbit_OP_GP_2020.tex}		%
 			\label{fig:OP_Qbit_SEGP}}
 		\subfloat[{Channel model \textit{Co-corr-I} with $\Tc = 200$.}]{\import{Figures/}{CDF_SE_OP_Case1_2020.tex}		%
 			\label{fig:OP_SE_GP_CoI}}
 		\\
 		\import{Figures/}{SE_OP_GP_legend.tex} 		
 		\\
 		\subfloat[{Channel model \textit{Cell-free} with $\Tc = 1000$.}]{\import{Figures/}{CDF_SE_OP_Case2_2020.tex}		%
 			\label{fig:OP_SE_GP_CoD1}}
 		\subfloat[{Channel model \textit{Cell-free}  with $\Tc = 200$.}]{\import{Figures/}{CDF_SE_OP_Case3_2020.tex}		%
 			\label{fig:OP_SE_GP_Cfree}}\\%
 	}
 	\caption{CDF of the optimal ADC bit allocation and SE per UE with $M=100$, $K = 10$, $\bar{\qul}/\sigma^2 = 1$ \mbox{(SNR $= 0$ [dB])}, and $b_\textsc{tot} = 3 M$. In Figs.~\ref{fig:OP_SE_GP_CoI},~\ref{fig:OP_SE_GP_CoD1},~and~\ref{fig:OP_SE_GP_Cfree},  the lines correspond to the results in Theorem~\ref{th:AchSE} and Corollary~\ref{cor:MR_SE_CL} (for MR) with the distortion model in \eqref{eq:distortion_e}. The markers are the results with exact quantization and integer bit allocation (see Section~\ref{sec:int_ADC}). The results for joint high/1-bit from \cite{Pirzadeh2018_Mixed_ADC}, assume 20 antennas with 11 bit ADCs and the rest with 1-bit ADCs.}
 	\label{fig:OP_HWI}
\hrulefill
 \end{figure*}
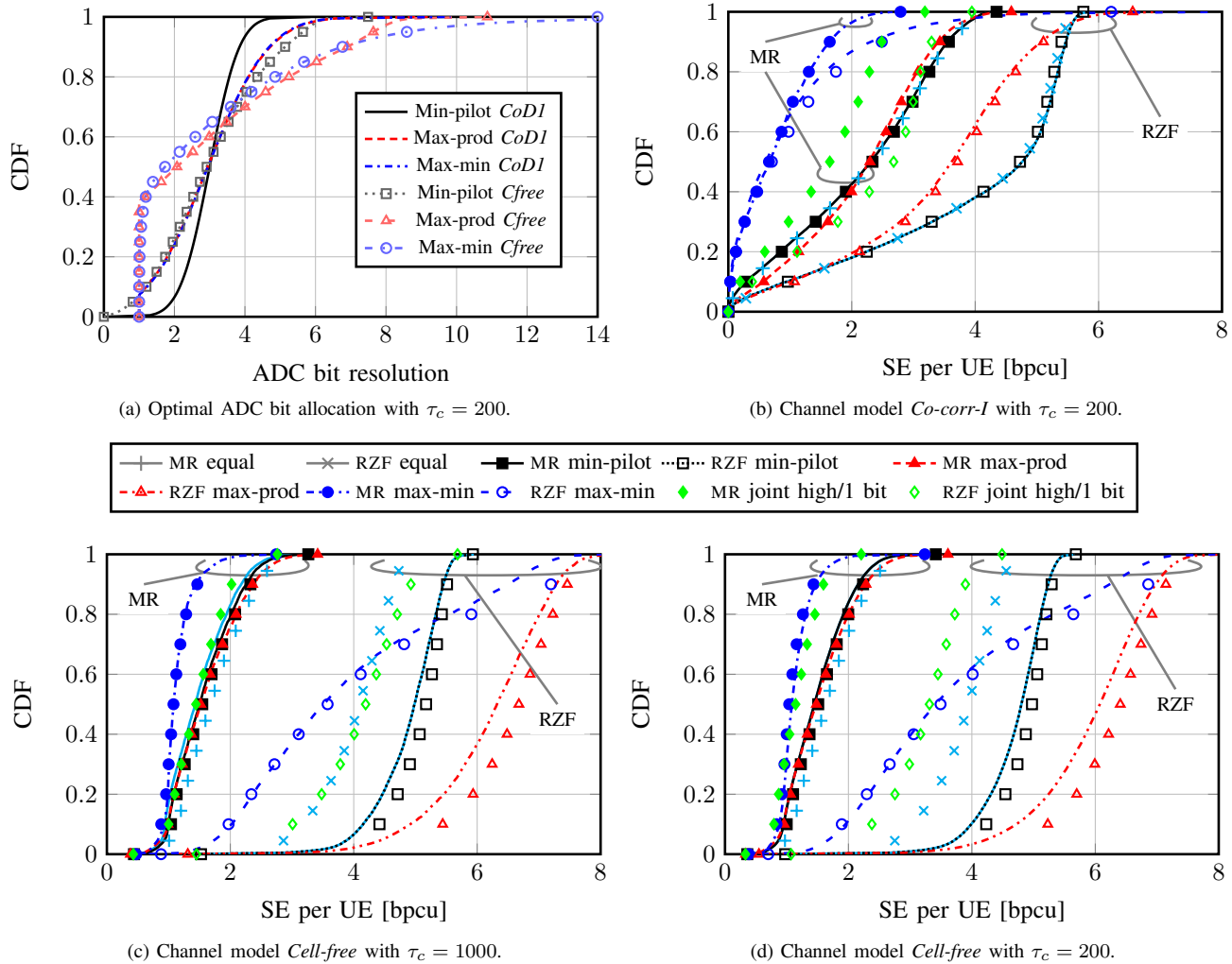

 To gain more insights into the SE with optimal HWI allocation, Monte-Carlo simulations are conducted following the same setup as in Section~\ref{subsec:sim_pilot_dist}. The optimization problems in \eqref{eq:OP_maxprod_fxPsi} and \eqref{eq:OP_maxmin-fxPsi} are solved with CVX, a package for specifying and solving convex programs \cite{grant_cvx_2014}. The iterative optimization algorithm converges in two iterations in all cases, thus, the convergence results are not plotted.

 Figure~\ref{fig:OP_Qbit_SEGP} shows the empirical CDF curve of the optimal ADC bit allocation for the \textit{Co-corr-D1} and \textit{Cell-free} cases. It can be seen that the optimal ADC bits are more spread in the case of \textit{Cell-free} since there are more LSF variations across the BS antennas. In the case when the bit allocation provides minimum pilot distortion, the ADC bits are less spread than for max-product of SINR or max-min fairness. In the minimum pilot distortion case, the optimal HWI allocation seeks to balance the dynamic range of the received pilot signal. In turn, when optimizing the effective SINR, more involved dependencies arise due to the linear combining and the optimal HWI allocation is conditioned on cross products between spatial correlation matrices.

Figs.~\ref{fig:OP_SE_GP_CoI},~\ref{fig:OP_SE_GP_CoD1},~and~\ref{fig:OP_SE_GP_Cfree}, show the empirical CDF of SE per UE for the additive uncorrelated distortion model in \eqref{eq:distortion_e} with the expressions found in Theorem~\ref{th:AchSE} for RZF and Corollary~\ref{cor:MR_SE_CL} for MR (depicted by lines). These results are validated by the exact quantization model introduced in Section~\ref{subsec:sim_pilot_dist} (depicted by the markers). The results with uniform equal ADC bit allocation, and the joint high-resolution/one-bit with antenna selection approach from \cite{Pirzadeh2018_Mixed_ADC} have been included as benchmarks. In the latter, the number of antennas with high-resolution ADCs is 20\% of the total number of BS antennas, and the value of the high ADC resolution is selected so that the total ADC bit budget constraint is satisfied, that is 11 bit ADCs. In Figs.~\ref{fig:OP_SE_GP_CoI},~\ref{fig:OP_SE_GP_CoD1},~and~\ref{fig:OP_SE_GP_Cfree}, it can be seen that max-min fairness reduces the SE of UEs with high channel gains (shown in the high percentile regions) in order to enhance the performance of the UE with minimum SE. Moreover, as more LSF variations arise going from \textit{Co-corr-I} to \textit{Cell-free}, max-min fairness is able to offer higher SE. However, except for very low SE percentile values, the max-product of SINR and minimum pilot distortion methods are able to offer higher SE. In the case of MR, max-product of SINR and minimum pilot distortion offer almost the same SE for all UEs.

 Notice that the optimization problems \eqref{eq:OP_maxprod_fxPsi} and \eqref{eq:OP_maxmin-fxPsi} are solved with the SINR expression for MR which does not correspond to the optimal operating point of RZF. Nevertheless, it is important to assess the potential benefits of interference suppression methods like RZF.
 The max-product of SINRs with MR tends to assign more power to UEs with higher channel gains which may not be the best approach to suppress interference. On the other hand, in cases with co-located BS antennas, the level of interference from a given UE varies less than in the \textit{Cell-free} case, where some BS antennas might be located much further than others. As a result, RZF with minimum pilot distortion provides higher SE for most UEs under \textit{Co-corr-I}, whereas the max-product of SINRs provides the highest SE under the \textit{Cell-free} case. 
 
The joint high-resolution/one-bit with antenna selection approach from \cite{Pirzadeh2018_Mixed_ADC} performs better in scenarios with large coherence block sizes as it is shown in Figure~\ref{fig:OP_SE_GP_CoD1}, achieving higher SE than max-min fairness with RZF for more than 60\% of the UEs. However, the minimum pilot distortion and max-product of SINRs methods offer higher SE than the joint high-resolution/one-bit with antenna selection approach. This indicates that allocating different ADC bit values based on LSF variations is more beneficial than having a few BS antennas equipped with high-resolution ADCs and the rest with one-bit ADCs.

In summary, statistical channel inversion power control combined with minimum pilot distortion HWI allocation offers the same or higher SE as the max-product of SINRs technique and it is easier to implement. The only exception is in the \textit{Cell-free} case with RZF, where optimizing the max-product of SINRs increases the SE for most UEs more than 1 [bpcu] compared to minimum pilot distortion, and more than 2 [bpcu] compared to equal ADC bit allocation.

\begin{remark}
		\label{rem:correlation_impact}				
		As shown in Figure~\ref{fig:OP_HWI}, unequal ADC bit allocation is mostly advantageous when there are LSF channel variations across the BS antennas, that is, large changes in the diagonal elements of the spatial correlation matrices. In contrast, the amount of correlation between antenna elements (i.e., value of the off-diagonal elements of the spatial correlation matrices) is less influential in the SE improvements provided by optimizing the ADC bit allocation. This result is in line with Remark~\ref{rem:min_pilot_dist}.
\end{remark}

\section{Impact of PwC on the HWI allocation}
\label{sec:PwC}
In previous sections, an ADC bit budget constraint is used to get a clear comparison between equal and mixed ADC bit allocation. However, in practice the PwC is a more suitable measure of cost. Thus, in this section the optimal HWI allocation for maximizing the product of SINRs under a PwC constraint is analyzed. This problem is solved based on a mixed-ADC approach, and to provide a benchmark for comparison, the solution to the same problem considering equal ADC bits across the BS antennas is also included.

Before formulating the optimization problem, a model for the PwC of ADCs needs to be established. However, finding a general power model for ADCs is hard because the PwC depends on many design parameters, for example, architecture, sampling frequency, oversampling factor, among others. Thus, to get a general model of the ADC PwC, the Walden's figure of merit  $\rm{FoM}_\textsc{w} = {\rm{P}_\textsc{adc}}/({2^b f_s})$ is used, where $\rm{P}_\textsc{adc}$ is the dissipated power by the ADC, $f_s$ is the sampling frequency, and $b$ is the number of ADC bits. Thus, the ADC PwC of the $m^{th}$ BS antenna is modeled as 
\begin{equation}
 {\rm{P}_\textsc{adc}}_m = \sf{D}_1  2^{b_m} = \sf{D}_1 \zeta_m \ep_m^{-1}
 \label{eq:ADCPwmodel}
\end{equation}
which grows exponentially with the number of ADC bits. The proportionality constant $\sf{D}_1$ depends on the Walden's figure of merit such that $\sf{D}_1 = \rm{FoM}_\textsc{W} f_s$. The survey \cite{B_Murmann_survey_2018} provides numerical values for these terms based on state-of-the-art ADCs.

Similar to the previous section, the optimization framework in \eqref{eq:general_OP} is used. Since the goal is to maximize the product of SINRs, the objective function is the same as in \eqref{eq:SINR_obj}. The key difference, is that in this analysis there is no fixed ADC bit budget, that is, the total ADC bit budget is optimized as well as the allocation of ADC bits. Thus, the constraint function is instead given by the PwC of the ADCs and the transmission data power, defined as 
\begin{align}
\label{eq:f_obj_minPw}
f^\textsc{hwi}(\bfs{\ep},\bfs{\pul}) &= \rm{P}_\textsc{txd-adc}(\bfs{\ep},\bfs{\pul}) - \gamma_\textsc{pc}\\
\rm{P}_\textsc{txd-adc}(\bfs{\ep},\bfs{\pul}) &= \underbrace{\left(1 - \frac{\Tp}{\Tc}\right)\frac{\Bw}{\eta}\sum_{k'=1}^{K}\pul_k}_{\text{Transmission data power}} + \underbrace{2 \sf{D}_1\sum_{m=1}^{M} \zeta_m \ep_m^{-1}}_{\text{Total ADC power}}
\label{eq:f_Ptx_adc}
\end{align}
where $\eta$ is the efficiency of the power amplifier, assumed to be equal for all BS antennas, and $\rm{P}_\textsc{txd-adc}(\bfs{\ep},\bfs{\pul})$ is the PwC that depends on the HWI level and transmission energy per data symbol. The term $\gamma_\textsc{pc}$ imposes a maximum PwC for $\rm{P}_\textsc{txd-adc}(\bfs{\ep},\bfs{\pul})$. The power function $\rm{P}_\textsc{txd-adc}(\bfs{\ep},\bfs{\pul})$ is linear in $\pul_k$ and monotonically decreasing in $\ep_m$. Thus, lower transmission power and higher HWI level result in lower PwC. Moreover, $\rm{P}_\textsc{txd-adc}(\bfs{\ep},\bfs{\pul})$ is a posynomial function, which means that the power constraint becomes a posynomial lower than a constant. The max-product of SINRs objective can be transformed into a monomial function by adding an auxiliary variable and including a geometric constraint as shown in Section~\ref{subsec:max_SINR}. Thus, the optimization problem in \eqref{eq:general_OP} with max-product of SINRs objective in \eqref{eq:SINR_obj} and constraint function in \eqref{eq:f_obj_minPw} is a geometric program \cite{Boyd_cvx_book}. In the case of equal ADC bit allocation, the optimization variable is changed to $\ep$ such that $\ep_m = \ep$ for $m\in\{1,\ldots,M\}$. Note that this change of variables also results in a geometric program.

To illustrate the balance between the sum SE benefit and the cost of PwC, the EE is used as a performance metric and is defined as \cite[Ch~5]{Bjornson2017_MaMIMObook}
 \begin{align*} 
\rm{EE} =& { \Bw \sum_{k=1}^{K}\rm{SE}_k}  \Bigg / \Bigg (\rm{P}_\textsc{cst} + \rm{P}_\textsc{ue}K + \rm{P}_\textsc{bs-a}M 
	\\&
	+ \rm{P}_\textsc{pilot}+ \rm{P}_\textsc{txd-adc}(\bfs{\ep},\bfs{\pul}) +   \rm{P}_\textsc{cd}\Bw \sum_{k=1}^{K}\rm{SE}_k \Bigg )
\stepcounter{equation}\tag{\theequation}
 \end{align*}
 where $\rm{P}_\textsc{cd}$ accounts for the coding, decoding and backhaul PwC. The transmission power of pilots is modeled by $\rm{P}_\textsc{pilot} = \frac{\Tp}{\Tc}\frac{\Bw}{\eta}\sum_{k'=1}^{K}\qul_k$. The circuit PwC per UE is $\rm{P}_\textsc{ue}$, while $\rm{P}_\textsc{bs-a}$ accounts for the power consumed by the circuitry on each BS antenna that is independent of the ADC bit resolution. The term $\rm{P}_\textsc{cst}$ accounts for the power consumed by baseband processing (e.g, linear processing and channel estimation) plus fixed terms (e.g., site cooling). The values of these parameters are in Table~\ref{tab:sim_param} and are selected based on \cite[Ch~5]{Bjornson2017_MaMIMObook}.%

\begin{figure*}[t]
	{
		\captionsetup{width=0.47\textwidth}
		\centering
		\hspace*{5mm}\import{Figures/}{Pw_OP_GP_legend.tex} 		
		\\
			\subfloat[CDF of optimal ADC bit with $\gamma_\textsc{pc}=10$ \textsc{[W]}.]{\import{Figures/}{CDF_Qbit_maxSE_tgPw_GP_2020.tex}
			\label{fig:CDF_Qbit_maxSE_tgPw}}
		\subfloat[Energy Efficiency vs $\gamma_\textsc{pc}$.]{\import{Figures/}{EE_vs_Pwtarget_GP.tex}		
			\label{fig:Pw_minPwGP_1}}
	}
	\caption{CDF of optimal ADC bit allocation and EE vs $\gamma_\textsc{pc}$ with RZF, $M=100$, $K=10$, $\Tc = 200$, $\bar{\qul} = \sigma^2$ \mbox{(SNR $= 0$ [dB])} and $\sf{D}_1 = 0.006$ \mbox{[W/conv-step]}. The results in Figure~\ref{fig:Pw_minPwGP_1} consider the exact quantization model and integer bit allocation (see Section~\ref{sec:int_ADC}). }
	\label{fig:OP_HWI_Pw}
\hrulefill
\end{figure*}
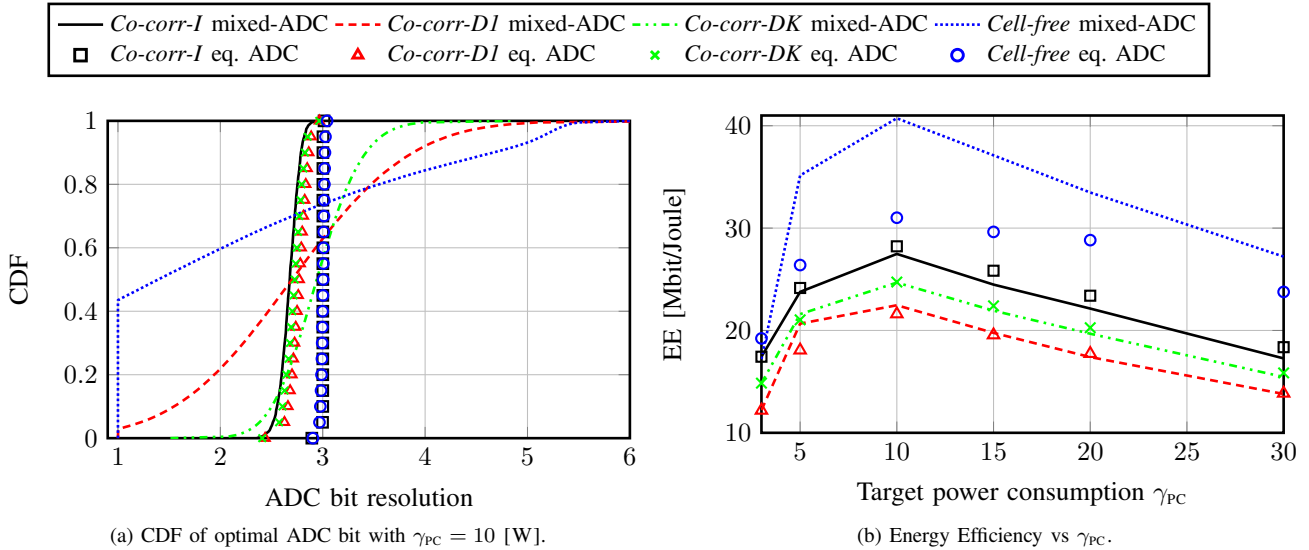

To analyze the optimal ADC bit allocation, Monte Carlo simulations with the same simulation setup as in Section~\ref{subsec:max_SINR} are performed. Figure~\ref{fig:CDF_Qbit_maxSE_tgPw} shows the empirical CDF of the optimal ADC bit allocation. It can be seen that, similar to Figure~\ref{fig:OP_Qbit_SEGP}, the ADC bit resolution is more spread as the variations of LSF among the channels increase. However, in contrast to Figure~\ref{fig:OP_Qbit_SEGP}, where the ADC resolutions reach values of up to 10 bits, the bit resolution in  Figure~\ref{fig:CDF_Qbit_maxSE_tgPw} is concentrated below 5 bits. In particular, in the case of \textit{Cell-free} with $\gamma_\textsc{pc} = 10$ [W], over 40 \% of the BS antennas have 1 bit ADC bit resolution. Thus, adding a PwC constraint reduces the values of the ADC bits used.

Figure~\ref{fig:Pw_minPwGP_1} shows the EE versus the target PwC, and  all results consider exact quantization with the proposed integer bit allocation in Section~\ref{sec:int_ADC} and RZF. Note that since the PwC and SE both increase when the ADC bit resolution increases, the optimal solution satisfies the power constraint with equality. Thus, the optimal SE follows the same behavior as the EE depicted in  Figure~\ref{fig:Pw_minPwGP_1}. The EE has unimodal shape (i.e., increases, saturates, and then decreases) which means that the target PwC can be customized to maximize SE and EE. Comparing equal and mixed-ADC bit allocations, it can be seen that for co-located scenarios, the EE is almost the same whereas in the \textit{Cell-free} case,
mixed-ADCs increases the EE up to 30\% at its maximum compared to equal ADC bit allocation.

\section{Conclusion}
\label{sec:conclusion}

This article studied the optimal ADC bit allocation problem for maximal SE or EE when the selection of ADC bits is done with respect to the large-scale fading. 
For the SE, in cases with co-located BS antennas and small channel variations across the antennas, equal ADC bit allocation is close to optimal. In contrast, when there are large channel variations across the BS antennas, as in cell-free Massive MIMO, the SE benefits of optimizing the ADC bit allocation are substantial for interference suppression methods such as RZF.
 
In the case of cell-free Massive MIMO, the proposed ADC bit allocation method to minimize the pilot distortion combined with statistical channel inversion power control and RZF has a simple implementation and provides more than 1 [bpcu] SE gain  compared to equal ADC bit allocation. Furthermore, for max-product of SINRs power control and ADC bit optimization, the SE gain is of 2 [bpcu] SE compared to equal ADC bit allocation.

By considering the PwC as a constraint in the max-product of SINRs problem, the optimal ADC bit allocation that maximizes EE and sum SE is found at ADC bit resolutions below 6 bits. Furthermore, in the case of cell-free Massive MIMO with RZF and a mixed-ADC approach, improvements of up to 30\% can be achieved compared to equal ADCs.

\begin{figure*}[t]
	\begin{align*}
	\bb{E}\left\{\bar{\bf{h}}_i^H\bf{D}_{\ep}^2\bf{D}_{h_i}\bar{\bf{h}}_i\right\} 
	&= \bb{E}\left\{\sum_{l=1}^{M}\sum_{m=1}^{M}\sum_{m'=1}^{M} [\bf{A}_k]_{ml} [\bf{A}_k]_{m'l}^* \ep_l^2	[\bf{R}_i^{\frac{1}{2}} \bf{g}]_m^*\left|[\bf{R}_i^{\frac{1}{2}}\bf{g}]_{l}\right|^2 [\bf{R}_i^{\frac{1}{2}}\bf{g}]_{m'} \right\}\\
	&\stackrel{\rm{(a_1)}}{=}
	\sum_{l=1}^{M}\ep_l^2 [\bf{R}_i\bf{A}_k ]_{ll} [\bf{R}_i\bf{A}_k ]_{ll}^* +  \sum_{l=1}^{M} \sum_{m=1}^{M}\ep_l^2  [\bf{A}_k]_{ml} [\bf{R}_i\bf{A}_k ]_{ml}^* [\bf{R}_i ]_{ll}
	\\
	&= \rm{tr}\left(\bf{D}_{\ep}^2\rm{abs}\left(\rm{diag}\left(\bf{R}_i\bf{R}_k\bfs{\Psi}_k^{-1} \right)\right)^2 + \bf{R}_i\bf{A}_k \bf{D}_{\ep}^2\bf{D}_{\bf{R}_{i}} \bf{A}_k^H\right)
    \tag{33}\label{eq:hDDh}
	\end{align*}	
		\hrulefill
	\begin{equation*}
	\bb{E}\left\{[\bf{g}]_m [\bf{g}]_{m'}^* [\bf{g}]_l [\bf{g}]_{l'}^*\right\}=\begin{cases}&\\%
	\bb{E}\left\{ \left|[\bf{g}]_m\right|^4\right\} = 2 & \text{ if }m=m'=l=l'\\%
	\bb{E}\left\{ \left|[\bf{g}]_m\right|^2 \left|[\bf{g}]_l\right|^2\right\} = 1 & \text{ if } m\neq l,\text{ with } \begin{cases}&\\%
	m=m',\, l=l'\\%
	m=l',\, l=m' \\%
	\end{cases}\\%
	0 & \text{ otherwise}.
	\end{cases}
	\tag{34}\label{eq:CN_prop}
	\end{equation*}
		\hrulefill
	\begin{align*}
	\bb{E}\left\{\rm{tr}\left( \bf{D}_{\ep}^2 \bf{D}_{h}^\textsc{d}\bf{A}_k \bf{D}_{\ep}^2\bf{D}_{h}^\textsc{p}\bf{A}_k^H\right)
	\right\} =& \rm{tr}\left( \bf{D}_{\ep}^2  \sum_{i=1}^{K}\pul_i\bf{D}_{\bf{R}_{i}} \bf{A}_k \bf{D}_{\ep}^2   \sum_{i'\neq i}^{K}\qul_{i'}\bf{D}_{\bf{R}_{i'}} \bf{A}_k^H\right)\\
	&
	+\bb{E}\left\{\sum_{i=1}^{K} \pul_i \qul_i \sum_{l=1}^{M}\sum_{m=1}^{M} \ep_{m}^2 \ep_{m'}^2 \left|[\bf{A}_k]_{m m'} \right|^2 \left|[\bf{R}_i^{\frac{1}{2}} \bf{g}]_m\right|^2 \left|[\bf{R}_i^{\frac{1}{2}} \bf{g}]_{m'}\right|^2   \right\}\\
	=&\rm{tr}\left( \bf{D}_{\ep}^2  \sum_{i=1}^{K}\pul_i\bf{D}_{\bf{R}_{i}} \bf{A}_k \bf{D}_{\ep}^2   \sum_{i'=1}^{K}\qul_{i'}\bf{D}_{\bf{R}_{i'}} \bf{A}_k^H\right)
	+ \sum_{i=1}^{K} \pul_i \qul_i   \bf{D}_{\ep}^2  \rm{diag}\left(\rm{abs}\left(\bf{A}_k\right)^2 \bf{D}_{\ep}^2  \rm{abs}\left(\bf{R}_i\right)^2 \right)
	\tag{35} \label{eq:ADhADh}
	\end{align*}

	\hrulefill
\end{figure*}

\appendices
\section*{Appendix A: Proof of Theorem~\ref{th:Pilot_dist}}
\manuallabel{app:proof_min_pilot_dist}{A}
Let $\lambda_1$ be the Lagrange multiplier for the constraint in \eqref{eq:OP_MSEChEst} and $\bfs{\ep}=[\ep_1,\ldots,\ep_M]^T$, then
\begin{equation}
\cal{L}(\bfs{\ep},\lambda_1) = \sum_{m=1}^{M}\ep_m^2p_m^\textsc{u} + \lambda_1\left(\sum_{m=1}^{M}\log_2\left(\frac{\zeta_m}{\ep_m}\right) - b_\textsc{tot}\right)
\end{equation}
is the Lagrangian function where $p_m^\textsc{u} = \sum_{i = 1}^{K}\qul_i[\bf{R}_i]_{mm}$ and the dual problem is
\begin{align*}
\begin{aligned}
& \maximize_{\lambda_1 \geq 0} & & \inf_{\ep_m \geq 0\;\forall m \in \{1,\ldots,M\}} \cal{L}(\bfs{\ep},\lambda_1).\\
\end{aligned}%
\stepcounter{equation}\label{eq:DualOP_MSEChEst}\tag{\theequation}
\end{align*}
Since \eqref{eq:OP_MSEChEst} is a convex problem, a point $(\hat{\bfs{\ep}},\,\hat{\lambda}_1)$ is primal and dual optimal (it solves \eqref{eq:OP_MSEChEst} and \eqref{eq:DualOP_MSEChEst}) with zero duality gap (i.e., the optimal objective is the same for \eqref{eq:OP_MSEChEst} and \eqref{eq:DualOP_MSEChEst}) when it satisfies the KKT conditions given by 
\begin{align*}
\stepcounter{equation}
\tag{\theequation.a} \label{eq:KKT_pilot_dist_a}
\vphantom{\left(\sum_{m=1}^{M}\log_2\left(\frac{\zeta_m}{\hat{\ep}_m}\right)\right)}\sum_{m=1}^{M}\log_2\left(\frac{\zeta_m}{\hat{\ep}_m}\right) - b_\textsc{tot} &\leq 0 
\\
\tag{\theequation.b} \label{eq:KKT_pilot_dist_b}
\hphantom{\sum_{m=1}^{M}\log_2\left(\frac{\zeta_m}{\hat{\ep}_m}\right) - -}\hat{\lambda}_1&\geq 0
\\
\tag{\theequation.c} \label{eq:KKT_pilot_dist_c}
\hat{\lambda}_1\left(\sum_{m=1}^{M}\log_2\left(\frac{\zeta_m}{\hat{\ep}_m}\right) - b_\textsc{tot}\right)  &= 0
\\
\tag{\theequation.d} \label{eq:KKT_pilot_dist_d}
\hphantom{\sum_{m=1}^{M}\log_2\left(\frac{\zeta_m}{\hat{\ep}_m}\right) \:}\nabla_{\bfs{\ep}}\cal{L}(\hat{\bfs{\ep}},\hat{\lambda}_1)  &= 0
\end{align*}
where $\nabla_{\bfs{\ep}}(\cdot)$ is the gradient with respect to $\bfs{\ep}$. From \eqref{eq:KKT_pilot_dist_d} the following conditions are found:
\begin{equation}
2\hat{\ep}_m p_m^\textsc{u} - \frac{\hat{\lambda}_1}{\ln(2) \hat{\ep}_m}=0,\;\forall m\in \{1,\ldots,M\}
\label{eq:KKT_pilot_dist_d1}
\end{equation}
where $\ln(\cdot)$ denotes the natural logarithm. By clearing $\hat{\ep}_m$ from \eqref{eq:KKT_pilot_dist_d1} and inserting it into \eqref{eq:KKT_pilot_dist_c}, the optimal solution is found and \eqref{eq:KKT_pilot_dist_a} is satisfied with equality.

\section*{Appendix B: Proof of closed-form SE expression with MR}
\manuallabel{app:proof_Ach_SE}{B}
The closed-form expression in \eqref{eq:SINR_MR_mat} is found by calculating the expectations in \eqref{eq:SINR_all_v} using known properties of circularly symmetric complex Gaussian random vectors \cite[Appx.~A]{Emil_scaling_laws}. First, the term in the numerator is found by using the fact that the LMMSE channel estimate $\hat{\bf{h}}_k$ and error $\tilde{\bf{h}}_k$ are uncorrelated by design, so that 
\begin{equation}
\bb{E}\left\{\bf{v}_k^H\bf{h}_k\right\} = \bb{E}\left\{\hat{\bf{h}}_k^H\left(\hat{\bf{h}}_k + \tilde{\bf{h}}_k\right)\right\} = \rm{tr}\left(\bf{R}_k\bfs{\Psi}_k^{-1}\bf{R}_k\right).
\label{eq:E_vHh_k}
\end{equation}
To calculate the first term in the denominator of \eqref{eq:SINR_all_v}, for ease of notation let $\bf{A}_k = \bf{R}_k\bfs{\Psi}_k^{-1}$, $\bar{\bf{h}}_i = \bf{A}_k^H\bf{h}_i$ and $\bf{D}_{h_i} = \rm{diag}(|[\bf{h}_i]_1|^2,\ldots, |[\bf{h}_i]_M|^2)$,  then
\begin{align*}
&\bb{E}\left\{\left|\bf{v}_k^H\bf{h}_i\right|^2\right\} 
=\bb{E}\left\{\vphantom{\frac{K}{K}}\right.
\left|\bar{\bf{h}}_i^H\bf{h}_k\right|^2 
+ \frac{\qul_i}{\Tp \qul_k} \bar{\bf{h}}_i^H\bf{D}_{\ep}^2\bf{D}_{h_i}\bar{\bf{h}}_i
\\&
 + \bar{\bf{h}}_i^H\left(\vphantom{\frac{K}{K}}\right. \frac{\sigma^2}{\Tp \qul_k}\bf{I}_M + \sum_{i'\neq i}^{K}\frac{\qul_{i'}}{\Tp \qul_k}\bf{D}_{\ep}^2\bf{D}_{\bf{R}_{i'}}   \left. \vphantom{\frac{K}{K}}\right)\bar{\bf{h}}_i\left. \vphantom{\frac{K}{K}}\right\}.
\stepcounter{equation}\tag{\theequation}\label{eq:abs2_v^Hh}
\end{align*}
From \cite[Appx.~A]{Emil_scaling_laws} it follows directly that 
\begin{equation}
\bb{E}\!\left\{\!\left|\bar{\bf{h}}_i^H\bf{h}_k\right|^{2}\!\right\} \!= \!\begin{cases}\!
\rm{tr}\left( \!\bf{A}_k \bf{R}_k \right)^2 \!+\! \rm{tr}\left(\!  \bf{R}_k\bf{A}_k \bf{R}_k\bf{A}_k^{H\,} \!\right) & \!\!\!\!\!\!\text{ for }i=k\\
\!\rm{tr}\left(\!  \bf{R}_i\bf{A}_k \bf{R}_k\bf{A}_k^{H\,} \!\right) & \!\!\!\!\!\!\text{ for }i\neq k.
\end{cases}
\end{equation}
To find the second term in \eqref{eq:abs2_v^Hh}, let $\bf{h}_i = \bf{R}_i^{\frac{1}{2}} \bf{g}$ such that $\bf{g}\sim\cal{CN}(\bf{0},\bf{I}_M)$ and $\bf{R}_i = \bf{R}_i^{\frac{1}{2}} \bf{R}_i^{\frac{1}{2}}$, then
the result in \eqref{eq:hDDh} at the top of this page holds,
where $\rm{(a_1)}$ follows from expanding the terms $[\bf{R}_i^{\frac{1}{2}}\bf{g}]_l = \sum_{n=1}^{M}[\bf{R}_i^{\frac{1}{2}}]_{m n} [\bf{g}]_n$ and applying the property of circularly symmetric complex Gaussian random vectors shown in \eqref{eq:CN_prop} at the top of this page.

The third term in \eqref{eq:abs2_v^Hh} is a quadratic function of $\bf{h}_i$ and the expectation is found by direct calculation of second order moments of  circularly symmetric complex Gaussian random vectors.
Since the noise $\bf{n}$ is independent from the channels and hardware distortions, the third term in the denominator of \eqref{eq:SINR_all_v} is
$\bb{E}\left\{ \left|\bf{v}_k^H\bf{n}\right|^2\right\} = \sigma^2\bb{E}\left\{ \|\hat{\bf{h}}_k \|^2\right\} = \sigma^2\rm{tr}\left(\bf{R}_k\bfs{\Psi}_k^{-1}\bf{R}_k\right)$.
To calculate the fourth term in the denominator of \eqref{eq:SINR_all_v}, recall that in the case of data transmission $\bf{D}_h = \sum_{i=1}^{K}\pul_i\bf{D}_{h_i}$ which in the following is denoted as $\bf{D}_h^\textsc{d}$, whereas in the case of pilot transmission $\bf{D}_h = \sum_{i=1}^{K}\qul_i\bf{D}_{h_i}$ which in the following is denoted as $\bf{D}_h^\textsc{p}$, then it follows that
\begin{align*}
\bb{E}\left\{\left|\bf{v}_k^H\bf{e}\right|^2\! \right\} \!=\!
 \bb{E}\!\Bigg\{ &\bar{\bf{h}}_k^H\bf{D}_{\ep}^2\bf{D}_{h}^\textsc{d}\bar{\bf{h}}_k \!+\! \frac{\sigma^2}{\Tp \qul_k}\rm{tr}\!\left( \bf{D}_{h}^\textsc{d}\bf{D}_{\ep}^2 \bf{A}_k \bf{A}_k^H\right)\!
 \\ & +\!\frac{1}{\Tp \qul_k}\rm{tr}\!\left( \bf{D}_{\ep}^2 \bf{D}_{h}^\textsc{d}\bf{A}_k \bf{D}_{\ep}^2\bf{D}_{h}^\textsc{p}\bf{A}_k^H\right)\!\!
 \Bigg\}\!.
\stepcounter{equation}\tag{\theequation} \label{eq:abs2_v^He}
\end{align*}
The first term in \eqref{eq:abs2_v^He} is follows as in \eqref{eq:hDDh}, and the second term in \eqref{eq:abs2_v^He} follows from calculating known second oder moments of the channel vectors. The last term in \eqref{eq:abs2_v^He} is computed as shown in \eqref{eq:ADhADh} at the top of this page,
where the last equality is found by expanding the terms $[\bf{R}_i^{\frac{1}{2}}\bf{g}]_l = \sum_{n=1}^{M}[\bf{R}_i^{\frac{1}{2}}]_{m n} [\bf{g}]_n$ and applying the properties in \eqref{eq:CN_prop}. Finally, by combining all results, merging terms that form the matrix $\bfs{\Psi}_k$ (see \eqref{eq:ChEst_Psi}), and performing lengthly algebraic manipulations, the expression in \eqref{eq:SINR_MR_mat} is found.

\bibliographystyle{IEEEtran}
\bibliography{Journal_References}

\end{document}

%% file: Figures/CDF_OpQbit_PilotDistK10.tex

\begin{tikzpicture}
	\begin{axis}[
	width=7cm,
	height=5.5cm,
	y label style = {at={(-0.03,0.5)},anchor=north},
	xlabel={Optimal ADC bit $b_m^\textsc{op}$},
	ylabel={CDF},
	xmin = 0,
	xmax = 6,
	ymin = 0,
	ymax = 1,
	xtick distance = 1,
	line width = 1pt,
	grid=both,
	legend columns = 1,
	legend style={at={(axis cs: 6,0.5)},font=\footnotesize,line width=1pt,draw=black,mark size=0.1pt},
	]	


	\addlegendentry{Co-corr-I}	
	\addlegendimage{black, mark=none, solid}
	\addlegendentry{Co-corr-D1}	
	\addlegendimage{red, mark=none,    densely dashed}
	\addlegendentry{Co-corr-DK}
	\addlegendimage{green, mark=none,    dashdotdotted}
	\addlegendentry{Cell-free}	
	\addlegendimage{blue, mark=none,    densely dotted}



	\addplot [black, mark=none, solid]
	table[x index={2}, y index={0}] {Figures/Prct_OptQbit.txt};									
	\addplot [red, mark=none,    densely dashed]
	table[x index={4}, y index={0}] {Figures/Prct_OptQbit.txt};			
	\addplot [green, mark=none,    dashdotdotted]
	table[x index={6}, y index={0}] {Figures/Prct_OptQbit.txt};			
	\addplot [blue, mark=none,    densely dotted]
	table[x index={8}, y index={0}] {Figures/Prct_OptQbit.txt};
%

	\end{axis}
	\end{tikzpicture}%

%% file: Figures/CDF_EqQbit_MR_2020.tex

\begin{tikzpicture}
	\begin{axis}[
	width=0.47\textwidth,
	height=0.32\textwidth,
	y label style = {at={(-0.03,0.5)},anchor=north},
	xlabel={SE per UE [bpcu]},
	ylabel={CDF},
	xmin = 0,
	xmax =5,
	ymin = 0,
	ymax = 1,
	line width = 1pt,
	grid=both,
	legend columns = 1,
	legend style={at={(axis cs: 5,0.55)},font=\footnotesize,line width=1pt,draw=black,mark size=0.1pt},
	]	


	\addlegendentry{Co-corr-I}	
	\addlegendimage{black, mark=square, solid,mark options={solid, line width = 0.8pt},mark size=2pt}	
	\addlegendentry{Co-corr-D1}	
	\addlegendimage{red, mark=triangle,    densely dashed,mark options={solid, line width = 0.8pt},mark size=2pt}
	\addlegendentry{Co-corr-DK}
	\addlegendimage{green, mark=x,    dashdotdotted,mark options={solid, line width = 0.9pt},mark size=2.3pt}
	\addlegendentry{Cell-free}	
	\addlegendimage{blue, mark=o,    densely dotted,mark options={solid, line width = 0.8pt},mark size=2pt}

	\addplot [black, mark=square, only marks,mark repeat=10,mark phase=1,mark options={solid, line width = 0.8pt},mark size=2pt]
	table[x index={1}, y index={0}] {Figures/CDF_SE_rev1_052020.txt};		

\addplot [black, mark=none, solid]
table[x index={65}, y index={0}] {Figures/CDF_SE_rev1_052020.txt};		
	\addplot [red, mark=triangle, only marks,mark repeat=10,mark phase=1,mark options={solid, line width = 0.8pt},mark size=2pt]
	table[x index={5}, y index={0}] {Figures/CDF_SE_rev1_052020.txt};			
					
	\addplot [red, mark=none,    densely dashed]
	table[x index={66}, y index={0}] {Figures/CDF_SE_rev1_052020.txt};			
	
	\addplot [green, mark=x, only marks,mark repeat=10,mark phase=1,mark options={solid, line width = 0.9pt},mark size=2.5pt]
	table[x index={9}, y index={0}] {Figures/CDF_SE_rev1_052020.txt};	
	\addplot [green, mark=none,    dashdotdotted]
	table[x index={67}, y index={0}] {Figures/CDF_SE_rev1_052020.txt};	

\addplot [blue, mark=o,only marks,mark repeat=10,mark phase=1,mark options={solid, line width = 0.8pt},mark size=2pt]
table[x index={13}, y index={0}] {Figures/CDF_SE_rev1_052020.txt};		

\addplot [blue, mark=none, densely dotted]
table[x index={68}, y index={0}] {Figures/CDF_SE_rev1_052020.txt};		

%

	\end{axis}
	\end{tikzpicture}%

%% file: Figures/CDF_EqQbit_RZF_2020.tex

\begin{tikzpicture}
	\begin{axis}[
	width=0.47\textwidth,
	height=0.32\textwidth,
	y label style = {at={(-0.03,0.5)},anchor=north},
	xlabel={SE per UE [bpcu]},
	ylabel={CDF},
	xmin = 0,
	xmax =6,
	ymin = 0,
	ymax = 1,
	line width = 1pt,
	grid=both,
	legend columns = 1,
	legend style={at={(axis cs: 2.5,0.95)},font=\footnotesize,line width=1pt,draw=black,mark size=0.1pt},
	]	


	\addlegendentry{Co-corr-I}	
	\addlegendimage{black, mark=square, solid,mark options={solid, line width = 0.8pt},mark size=2pt}	
	\addlegendentry{Co-corr-D1}	
	\addlegendimage{red, mark=triangle,    densely dashed,mark options={solid, line width = 0.8pt},mark size=2pt}
	\addlegendentry{Co-corr-DK}
	\addlegendimage{green, mark=x,    dashdotdotted,mark options={solid, line width = 0.9pt},mark size=2.3pt}
	\addlegendentry{Cell-free}	
	\addlegendimage{blue, mark=o,    densely dotted,mark options={solid, line width = 0.8pt},mark size=2pt}

	\addplot [black, mark=square, only marks,mark repeat=10,mark phase=1,mark options={solid, line width = 0.8pt},mark size=2pt]
	table[x index={2}, y index={0}] {Figures/CDF_SE_rev1_052020.txt};		

\addplot [black, mark=none, solid]
table[x index={4}, y index={0}] {Figures/CDF_SE_rev1_052020.txt};		
				
	\addplot [red, mark=triangle, only marks,mark repeat=10,mark phase=1,mark options={solid, line width = 0.8pt},mark size=2pt]
	table[x index={6}, y index={0}] {Figures/CDF_SE_rev1_052020.txt};			

	\addplot [red, mark=none,    densely dashed]
	table[x index={8}, y index={0}] {Figures/CDF_SE_rev1_052020.txt};

	\addplot [green, mark=x, only marks,mark repeat=10,mark phase=1,mark options={solid, line width = 0.9pt},mark size=2.5pt]
	table[x index={10}, y index={0}] {Figures/CDF_SE_rev1_052020.txt};	
	\addplot [green, mark=none,    dashdotdotted]
	table[x index={12}, y index={0}] {Figures/CDF_SE_rev1_052020.txt};

\addplot [blue, mark=o,only marks,mark repeat=10,mark phase=1,mark options={solid, line width = 0.8pt},mark size=2pt]
table[x index={14}, y index={0}] {Figures/CDF_SE_rev1_052020.txt};		

\addplot [blue, mark=none, densely dotted]
table[x index={16}, y index={0}] {Figures/CDF_SE_rev1_052020.txt};		

	\end{axis}
	\end{tikzpicture}%

%% file: Figures/CDF_minPilotQbit_MR_2020.tex

\begin{tikzpicture}
	\begin{axis}[
	width=0.47\textwidth,
	height=0.32\textwidth,
	y label style = {at={(-0.03,0.5)},anchor=north},
	xlabel={SE per UE [bpcu]},
	ylabel={CDF},
	xmin = 0,
	xmax =5,
	ymin = 0,
	ymax = 1,
	line width = 1pt,
	grid=both,
	legend columns = 1,
	legend style={at={(axis cs: 5,0.55)},font=\footnotesize,line width=1pt,draw=black,mark size=0.1pt},
	]	


	\addlegendentry{Co-corr-I}	
	\addlegendimage{black, mark=square, solid,mark options={solid, line width = 0.8pt},mark size=2pt}	
	\addlegendentry{Co-corr-D1}	
	\addlegendimage{red, mark=triangle,    densely dashed,mark options={solid, line width = 0.8pt},mark size=2pt}
	\addlegendentry{Co-corr-DK}
	\addlegendimage{green, mark=x,    dashdotdotted,mark options={solid, line width = 0.9pt},mark size=2.3pt}
	\addlegendentry{Cell-free}	
	\addlegendimage{blue, mark=o,    densely dotted,mark options={solid, line width = 0.8pt},mark size=2pt}

	\addplot [black, mark=square, only marks,mark repeat=10,mark phase=1,mark options={solid, line width = 0.8pt},mark size=2pt]
	table[x index={17}, y index={0}] {Figures/CDF_SE_rev1_052020.txt};		

\addplot [black, mark=none, solid]
table[x index={69}, y index={0}] {Figures/CDF_SE_rev1_052020.txt};		
	\addplot [red, mark=triangle, only marks,mark repeat=10,mark phase=1,mark options={solid, line width = 0.8pt},mark size=2pt]
	table[x index={21}, y index={0}] {Figures/CDF_SE_rev1_052020.txt};			

	\addplot [red, mark=none,    densely dashed]
	table[x index={70}, y index={0}] {Figures/CDF_SE_rev1_052020.txt};			

	\addplot [green, mark=x, only marks,mark repeat=10,mark phase=1,mark options={solid, line width = 0.9pt},mark size=2.5pt]
	table[x index={25}, y index={0}] {Figures/CDF_SE_rev1_052020.txt};	
	\addplot [green, mark=none,    dashdotdotted]
	table[x index={71}, y index={0}] {Figures/CDF_SE_rev1_052020.txt};	

\addplot [blue, mark=o,only marks,mark repeat=10,mark phase=1,mark options={solid, line width = 0.8pt},mark size=2pt]
table[x index={29}, y index={0}] {Figures/CDF_SE_rev1_052020.txt};		

\addplot [blue, mark=none, densely dotted]
table[x index={72}, y index={0}] {Figures/CDF_SE_rev1_052020.txt};		

	\end{axis}
	\end{tikzpicture}%

%% file: Figures/CDF_minPilotQbit_RZF_2020.tex

\begin{tikzpicture}
	\begin{axis}[
	width=0.47\textwidth,
	height=0.32\textwidth,
	y label style = {at={(-0.03,0.5)},anchor=north},
	xlabel={SE per UE [bpcu]},
	ylabel={CDF},
	xmin = 0,
	xmax =6,
	ymin = 0,
	ymax = 1,
	line width = 1pt,
	grid=both,
	legend columns = 1,
	legend style={at={(axis cs: 2.5,0.95)},font=\footnotesize,line width=1pt,draw=black,mark size=0.1pt},
	]	


	\addlegendentry{Co-corr-I}	
	\addlegendimage{black, mark=square, solid,mark options={solid, line width = 0.8pt},mark size=2pt}	
	\addlegendentry{Co-corr-D1}	
	\addlegendimage{red, mark=triangle,    densely dashed,mark options={solid, line width = 0.8pt},mark size=2pt}
	\addlegendentry{Co-corr-DK}
	\addlegendimage{green, mark=x,    dashdotdotted,mark options={solid, line width = 0.9pt},mark size=2.3pt}
	\addlegendentry{Cell-free}	
	\addlegendimage{blue, mark=o,    densely dotted,mark options={solid, line width = 0.8pt},mark size=2pt}

	\addplot [black, mark=square, only marks,mark repeat=10,mark phase=1,mark options={solid, line width = 0.8pt},mark size=2pt]
	table[x index={18}, y index={0}] {Figures/CDF_SE_rev1_052020.txt};		

\addplot [black, mark=none, solid]
table[x index={20}, y index={0}] {Figures/CDF_SE_rev1_052020.txt};		
				
	\addplot [red, mark=triangle, only marks,mark repeat=10,mark phase=1,mark options={solid, line width = 0.8pt},mark size=2pt]
	table[x index={22}, y index={0}] {Figures/CDF_SE_rev1_052020.txt};			

	\addplot [red, mark=none,    densely dashed]
	table[x index={24}, y index={0}] {Figures/CDF_SE_rev1_052020.txt};

	\addplot [green, mark=x, only marks,mark repeat=10,mark phase=1,mark options={solid, line width = 0.9pt},mark size=2.5pt]
	table[x index={26}, y index={0}] {Figures/CDF_SE_rev1_052020.txt};	
	\addplot [green, mark=none,    dashdotdotted]
	table[x index={28}, y index={0}] {Figures/CDF_SE_rev1_052020.txt};

\addplot [blue, mark=o,only marks,mark repeat=10,mark phase=1,mark options={solid, line width = 0.8pt},mark size=2pt]
table[x index={30}, y index={0}] {Figures/CDF_SE_rev1_052020.txt};		

\addplot [blue, mark=none, densely dotted]
table[x index={32}, y index={0}] {Figures/CDF_SE_rev1_052020.txt};		

	\end{axis}
	\end{tikzpicture}%

%% file: Figures/CDF_Qbit_OP_GP_2020.tex

\begin{tikzpicture}
	\begin{axis}[
	width=0.47\textwidth,
	height=0.32\textwidth,
	y label style = {at={(-0.03,0.5)},anchor=north},
	xlabel={ADC bit resolution},
	ylabel={CDF},
	xmin = 0,
	xmax =14,
	ymin = 0,
	ymax = 1,
	line width = 1pt,
	grid=both,
	legend columns = 1,
	legend style={at={(axis cs: 13,0.75)},font=\footnotesize,line width=1pt,draw=black,mark size=0.1pt},
	]	


	\addlegendentry{Min-pilot \textit{CoD1}}	
	\addlegendimage{black, mark=none, solid,mark options={solid, line width = 0.8pt},mark size=2pt}	
	\addlegendentry{Max-prod \textit{CoD1}}	
	\addlegendimage{red, mark=none,    densely dashed,mark options={solid, line width = 0.8pt},mark size=2pt}
	\addlegendentry{Max-min \textit{CoD1}}
	\addlegendimage{blue, mark=none,    dashdotted,mark options={solid, line width = 0.8pt},mark size=2pt}
	
	\addlegendentry{Min-pilot \textit{Cfree}}
	\addlegendimage{black!60!white, mark=square, dotted,mark options={solid, line width = 0.8pt},mark size=1.8pt}	
		\addlegendentry{Max-prod \textit{Cfree}}	
	\addlegendimage{red!60!white, mark=triangle,    loosely dashed,mark options={solid, line width = 0.8pt},mark size=2pt}
	\addlegendentry{Max-min \textit{Cfree}}
	\addlegendimage{blue!60!white, mark=o,  loosely  dashdotted,mark options={solid, line width = 0.8pt},mark size=2pt}
	
\addplot [black, mark=none, solid]
table[x index={12}, y index={0}] {Figures/CDF_Qbit_rev1_052020.txt};						
\addplot [black!60!white, mark=square, dotted,mark repeat=10,mark phase=1,mark options={solid, line width = 0.8pt},mark size=1.6pt]
table[x index={16}, y index={0}] {Figures/CDF_Qbit_rev1_052020.txt};


\addplot [red, mark=none,    densely dashed]
table[x index={20}, y index={0}] {Figures/CDF_Qbit_rev1_052020.txt};			
\addplot [red!60!white, mark=triangle,     dashed,mark repeat=10,mark phase=1,mark options={solid, line width = 0.8pt},mark size=2pt]
table[x index={24}, y index={0}] {Figures/CDF_Qbit_rev1_052020.txt};

	\addplot [blue, mark=none,    dashdotted]
	table[x index={28}, y index={0}] {Figures/CDF_Qbit_rev1_052020.txt};	
	\addplot [blue!60!white, mark=o,   loosely dashdotted,mark repeat=10,mark phase=1,mark options={solid, line width = 0.8pt},mark size=2pt]
	table[x index={32}, y index={0}] {Figures/CDF_Qbit_rev1_052020.txt};

%
%
%

	\end{axis}
	\end{tikzpicture}%

	

%% file: Figures/CDF_SE_OP_Case1_2020.tex

\begin{tikzpicture}
	\begin{axis}[
	width=0.47\textwidth,
	height=0.32\textwidth,
	y label style = {at={(-0.03,0.5)},anchor=north},
	xlabel={SE per UE [bpcu]},
	ylabel={CDF},
	xmin = 0,
	xmax =8,
	ymin = 0,
	ymax = 1,
	line width = 1pt,
	grid=both,
	]	


%


\draw (axis cs:0.6,0.85) node[draw=none,fill =white,align = center] {\footnotesize{MR}};
\draw[-,color = gray] (axis cs:0.6,0.9) -- (axis cs:1.8,0.96);	
\draw[-,color = gray] plot [smooth , tension=3.2 ] coordinates{ (axis cs: 1.85,0.98) (axis cs: 2.05,0.95) (axis cs: 2.3,0.98)   };

\draw[-,color = gray] (axis cs:0.6,0.8) -- (axis cs:1.5,0.48);	
\draw[-,color = gray] plot [smooth , tension=3 ] coordinates{ (axis cs: 1.5,0.48) (axis cs: 1.9,0.43) (axis cs: 2.3,0.48)   };

\draw (axis cs:7,0.6	) node[draw=none,fill =white,align = center] {\footnotesize{RZF}};
\draw[-,color = gray] (axis cs:7,0.65) -- (axis cs:6.2,0.94);	
\draw[-,color = gray] plot [smooth , tension=3 ] coordinates{ (axis cs: 5,0.98) (axis cs: 5.6,0.93) (axis cs: 6.2,0.98)   };


\addplot [cyan, mark=none, line width = 0.8pt, solid]
table[x index={65}, y index={0}] {Figures/CDF_SE_rev1_052020.txt};		

\addplot [cyan, mark=none, line width = 1.0pt, solid]
table[x index={4}, y index={0}] {Figures/CDF_SE_rev1_052020.txt};

	\addplot [black, mark=square*, only marks,mark repeat=20,mark phase=0,mark options={solid, line width = 0.8pt},mark size=2pt]
	table[x index={17}, y index={0}] {Figures/CDF_SE_rev1_052020.txt};		
	
	\addplot [black, mark=none, solid]
	table[x index={69}, y index={0}] {Figures/CDF_SE_rev1_052020.txt};						

	\addplot [red, mark=triangle*, only marks,mark repeat=20,mark phase=0,mark options={solid, line width = 0.8pt},mark size=2pt]
	table[x index={33}, y index={0}] {Figures/CDF_SE_rev1_052020.txt};

	\addplot [red, mark=none,    densely dashed]
	table[x index={73}, y index={0}] {Figures/CDF_SE_rev1_052020.txt};			
	\addplot [blue, mark=*, only marks,mark repeat=20,mark phase=0,mark options={solid, line width = 0.8pt},mark size=2pt]
	table[x index={49}, y index={0}] {Figures/CDF_SE_rev1_052020.txt};	
	
	\addplot [blue, mark=none,    dashdotted]
	table[x index={77}, y index={0}] {Figures/CDF_SE_rev1_052020.txt};	
					35

	
	\addplot [black, mark=square, only marks,mark repeat=20,mark phase=0,mark options={solid, line width = 0.8pt},mark size=2pt]
	table[x index={18}, y index={0}] {Figures/CDF_SE_rev1_052020.txt};		
	
	\addplot [black, mark=none, densely dotted]
	table[x index={20}, y index={0}] {Figures/CDF_SE_rev1_052020.txt};

	
	\addplot [red, mark=triangle, only marks,mark repeat=20,mark phase=0,mark options={solid, line width = 0.8pt},mark size=2pt]
	table[x index={34}, y index={0}] {Figures/CDF_SE_rev1_052020.txt};

	\addplot [red, mark=none,    dashdotdotted]
	table[x index={36}, y index={0}] {Figures/CDF_SE_rev1_052020.txt};			
	
	\addplot [blue, mark=o, only marks,mark repeat=20,mark phase=0,mark options={solid, line width = 0.8pt},mark size=2pt]
	table[x index={50}, y index={0}] {Figures/CDF_SE_rev1_052020.txt};	
	
	\addplot [blue, mark=none,  loosely  dashed]
	table[x index={52}, y index={0}] {Figures/CDF_SE_rev1_052020.txt};	
	
	
	
\addplot [green, mark=diamond*, only marks,mark repeat=20,mark phase=0,mark options={solid, line width = 0.8pt},mark size=2pt]
table[x index={17}, y index={0}] {Figures/CDF_SE_mxHigh1Qbit_tc200_062020.txt};

\addplot [green, mark=diamond, only marks,mark repeat=20,mark phase=0,mark options={solid, line width = 0.8pt},mark size=2pt]
table[x index={18}, y index={0}] {Figures/CDF_SE_mxHigh1Qbit_tc200_062020.txt};	


\addplot [cyan, mark=+,only marks,mark repeat=20,mark phase=10,mark options={solid, line width = 0.8pt},mark size=2.8pt]
table[x index={1}, y index={0}] {Figures/CDF_SE_rev1_052020.txt};		

\addplot [cyan, mark=x,only marks,mark repeat=20,mark phase=10,mark options={solid, line width = 0.8pt},mark size=2.8pt]
table[x index={2}, y index={0}] {Figures/CDF_SE_rev1_052020.txt};


	\end{axis}
	\end{tikzpicture}%

%% file: Figures/SE_OP_GP_legend.tex

\begin{tikzpicture}
	\begin{axis}[
	width=0.8\textwidth,
	height=50pt,
	axis line style={draw=none},
	tick style={draw=none},
	xticklabels=\empty,
	yticklabels=\empty,
	xmin = 0,
	xmax = 500,
	ymin = 0,
	ymax = 1,
	line width = 1pt,
	legend columns = 5,
	legend cell align={left},
	legend style={at={(0.5,0)},anchor=north,font=\small,line width=1pt,draw=black,mark size=0.1pt},
	]

		\addlegendentry{\textsc{mr} equal}	
		\addlegendimage{gray, mark=+, solid,mark options={solid, line width = 0.8pt},mark size=3pt}					
		\addlegendentry{\textsc{rzf} equal}	
		\addlegendimage{gray, mark=x, solid,mark options={solid, line width = 0.8pt},mark size=3pt}		
		
		\addlegendentry{\textsc{mr} min-pilot}	
		\addlegendimage{black, mark=square*, solid,mark options={solid, line width = 0.8pt},mark size=2pt}			
		\addlegendentry{\textsc{rzf} min-pilot}	
		\addlegendimage{black, mark=square, densely dotted,mark options={solid, line width = 0.8pt},mark size=2pt}		
		
		\addlegendentry{\textsc{mr} max-prod}	
		\addlegendimage{red, mark=triangle*,    densely dashed,mark options={solid, line width = 0.8pt},mark size=2.2pt}		
		\addlegendentry{\textsc{rzf} max-prod}
		\addlegendimage{red, mark=triangle,    dashdotdotted,mark options={solid, line width = 0.8pt},mark size=2.2pt}
		
		\addlegendentry{\textsc{mr} max-min}	
		\addlegendimage{blue, mark=*,    dashdotted,mark options={solid, line width = 0.8pt},mark size=2pt}			
		\addlegendentry{\textsc{rzf} max-min}	
		\addlegendimage{blue, mark=o,    loosely  dashed,mark options={solid, line width = 0.8pt},mark size=2pt}

		\addlegendentry{\textsc{mr} joint high/1 bit}	
		\addlegendimage{green, mark=diamond*,    only marks,mark options={solid, line width = 0.8pt},mark size=2.2pt}			
		\addlegendentry{\textsc{rzf} joint high/1 bit}	
		\addlegendimage{green, mark=diamond,    only marks,mark options={solid, line width = 0.8pt},mark size=2.2pt}

%
%
%

	\end{axis}
	\end{tikzpicture}%

%% file: Figures/CDF_SE_OP_Case2_2020.tex

\begin{tikzpicture}
	\begin{axis}[
	width=0.47\textwidth,
	height=0.32\textwidth,
	y label style = {at={(-0.03,0.5)},anchor=north},
	xlabel={SE per UE [bpcu]},
	ylabel={CDF},
	xmin = 0,
	xmax =8,
	ymin = 0,
	ymax = 1,
	line width = 1pt,
	grid=both,
	]	


%


\draw (axis cs:0.6,0.85) node[draw=none,fill =white,align = center] {\footnotesize{MR}};
\draw[-,color = gray] (axis cs:0.6,0.9) -- (axis cs:1.6,0.94);	
\draw[-,color = gray] plot [smooth , tension=3 ] coordinates{ (axis cs: 1.6,0.98) (axis cs: 2.3,0.93) (axis cs: 3.2,0.98)   };

\draw (axis cs:7.3,0.45	) node[draw=none,fill =white,align = center] {\footnotesize{RZF}};
\draw[-,color = gray] (axis cs:7.3,0.5) -- (axis cs:5.8,0.93);	
\draw[-,color = gray] plot [smooth , tension=3 ] coordinates{ (axis cs: 4.5,0.98) (axis cs: 6.15,0.93) (axis cs: 7.8,0.98)   };


	\addplot [cyan, mark=none, line width = 1.0pt, solid]
	table[x index={84}, y index={0}] {Figures/CDF_SE_mxHigh1Qbit_tc1000_062020.txt};		
	\addplot [cyan, mark=+,only marks,mark repeat=20,mark phase=10,mark options={solid, line width = 0.8pt},mark size=2.5pt]
	table[x index={13}, y index={0}] {Figures/CDF_SE_mxHigh1Qbit_tc1000_062020.txt};		
%
	\addplot [cyan, mark=none, line width = 1.0pt, solid]
	table[x index={16}, y index={0}] {Figures/CDF_SE_mxHigh1Qbit_tc1000_062020.txt};
	\addplot [cyan, mark=x,only marks,mark repeat=20,mark phase=10,mark options={solid, line width = 0.8pt},mark size=2.5pt]
	table[x index={14}, y index={0}] {Figures/CDF_SE_mxHigh1Qbit_tc1000_062020.txt};

	\addplot [black, mark=square*, only marks,mark repeat=20,mark phase=0,mark options={solid, line width = 0.8pt},mark size=2pt]
	table[x index={29}, y index={0}] {Figures/CDF_SE_mxHigh1Qbit_tc1000_062020.txt};		
	
	\addplot [black, mark=none, solid]
	table[x index={88}, y index={0}] {Figures/CDF_SE_mxHigh1Qbit_tc1000_062020.txt};						

	\addplot [red, mark=triangle*, only marks,mark repeat=20,mark phase=0,mark options={solid, line width = 0.8pt},mark size=2pt]
	table[x index={45}, y index={0}] {Figures/CDF_SE_mxHigh1Qbit_tc1000_062020.txt};

	\addplot [red, mark=none,    densely dashed]
	table[x index={92}, y index={0}] {Figures/CDF_SE_mxHigh1Qbit_tc1000_062020.txt};			

	\addplot [blue, mark=*, only marks,mark repeat=20,mark phase=0,mark options={solid, line width = 0.8pt},mark size=2pt]
	table[x index={61}, y index={0}] {Figures/CDF_SE_mxHigh1Qbit_tc1000_062020.txt};	
	
	\addplot [blue, mark=none,    dashdotted]
	table[x index={96}, y index={0}] {Figures/CDF_SE_mxHigh1Qbit_tc1000_062020.txt};	
					%


	\addplot [black, mark=square, only marks,mark repeat=20,mark phase=0,mark options={solid, line width = 0.8pt},mark size=2pt]
	table[x index={30}, y index={0}] {Figures/CDF_SE_mxHigh1Qbit_tc1000_062020.txt};		
	
	\addplot [black, mark=none, densely dotted]
	table[x index={32}, y index={0}] {Figures/CDF_SE_mxHigh1Qbit_tc1000_062020.txt};

	\addplot [red, mark=triangle, only marks,mark repeat=20,mark phase=0,mark options={solid, line width = 0.8pt},mark size=2pt]
	table[x index={46}, y index={0}] {Figures/CDF_SE_mxHigh1Qbit_tc1000_062020.txt};

	\addplot [red, mark=none,    dashdotdotted]
	table[x index={48}, y index={0}] {Figures/CDF_SE_mxHigh1Qbit_tc1000_062020.txt};

	\addplot [blue, mark=o, only marks,mark repeat=20,mark phase=0,mark options={solid, line width = 0.8pt},mark size=2pt]
	table[x index={62}, y index={0}] {Figures/CDF_SE_mxHigh1Qbit_tc1000_062020.txt};	
	
	\addplot [blue, mark=none,  loosely  dashed]
	table[x index={64}, y index={0}] {Figures/CDF_SE_mxHigh1Qbit_tc1000_062020.txt};


%
\addplot [green, mark=diamond*, only marks,mark repeat=20,mark phase=0,mark options={solid, line width = 0.8pt},mark size=2pt]
table[x index={77}, y index={0}] {Figures/CDF_SE_mxHigh1Qbit_tc1000_062020.txt};

\addplot [green, mark=diamond, only marks,mark repeat=20,mark phase=0,mark options={solid, line width = 0.8pt},mark size=2pt]
table[x index={78}, y index={0}] {Figures/CDF_SE_mxHigh1Qbit_tc1000_062020.txt};

	\end{axis}
	\end{tikzpicture}%

%% file: Figures/CDF_SE_OP_Case3_2020.tex

\begin{tikzpicture}
	\begin{axis}[
	width=0.47\textwidth,
	height=0.32\textwidth,
	y label style = {at={(-0.03,0.5)},anchor=north},
	xlabel={SE per UE [bpcu]},
	ylabel={CDF},
	xmin = 0,
	xmax =8,
	ymin = 0,
	ymax = 1,
	line width = 1pt,
	grid=both,
	]	


%


\draw (axis cs:0.6,0.85) node[draw=none,fill =white,align = center] {\footnotesize{MR}};
\draw[-,color = gray] (axis cs:0.6,0.9) -- (axis cs:1.5,0.93);	
\draw[-,color = gray] plot [smooth , tension=3 ] coordinates{ (axis cs: 1.5,0.98) (axis cs: 2.35,0.93) (axis cs: 3.2,0.98)   };

\draw (axis cs:7.3,0.5	) node[draw=none,fill =white,align = center] {\footnotesize{RZF}};
\draw[-,color = gray] (axis cs:7.3,0.55) -- (axis cs:6.2,0.93);	
\draw[-,color = gray] plot [smooth , tension=3 ] coordinates{ (axis cs: 4.2,0.98) (axis cs: 5.85,0.93) (axis cs: 7.5,0.98)   };


\addplot [cyan, mark=none, line width = 1.0pt, solid]
table[x index={15}, y index={0}] {Figures/CDF_SE_rev1_052020.txt};		
\addplot [cyan, mark=+,only marks,mark repeat=20,mark phase=10,mark options={solid, line width = 0.8pt},mark size=2.5pt]
table[x index={13}, y index={0}] {Figures/CDF_SE_rev1_052020.txt};		

\addplot [cyan, mark=none, line width = 1.0pt, solid]
table[x index={16}, y index={0}] {Figures/CDF_SE_rev1_052020.txt};
\addplot [cyan, mark=x,only marks,mark repeat=20,mark phase=10,mark options={solid, line width = 0.8pt},mark size=2.5pt]
table[x index={14}, y index={0}] {Figures/CDF_SE_rev1_052020.txt};

	\addplot [black, mark=square*, only marks,mark repeat=20,mark phase=0,mark options={solid, line width = 0.8pt},mark size=2pt]
	table[x index={29}, y index={0}] {Figures/CDF_SE_rev1_052020.txt};		
	
	\addplot [black, mark=none, solid]
	table[x index={31}, y index={0}] {Figures/CDF_SE_rev1_052020.txt};						

	\addplot [red, mark=triangle*, only marks,mark repeat=20,mark phase=0,mark options={solid, line width = 0.8pt},mark size=2pt]
	table[x index={45}, y index={0}] {Figures/CDF_SE_rev1_052020.txt};

	\addplot [red, mark=none,    densely dashed]
	table[x index={47}, y index={0}] {Figures/CDF_SE_rev1_052020.txt};			

	\addplot [blue, mark=*, only marks,mark repeat=20,mark phase=0,mark options={solid, line width = 0.8pt},mark size=2pt]
	table[x index={61}, y index={0}] {Figures/CDF_SE_rev1_052020.txt};	
	
	\addplot [blue, mark=none,    dashdotted]
	table[x index={63}, y index={0}] {Figures/CDF_SE_rev1_052020.txt};	
	

	\addplot [black, mark=square, only marks,mark repeat=20,mark phase=0,mark options={solid, line width = 0.8pt},mark size=2pt]
	table[x index={30}, y index={0}] {Figures/CDF_SE_rev1_052020.txt};		
	
	\addplot [black, mark=none, densely dotted]
	table[x index={32}, y index={0}] {Figures/CDF_SE_rev1_052020.txt};

	\addplot [red, mark=triangle, only marks,mark repeat=20,mark phase=0,mark options={solid, line width = 0.8pt},mark size=2pt]
	table[x index={46}, y index={0}] {Figures/CDF_SE_rev1_052020.txt};

	\addplot [red, mark=none,    dashdotdotted]
	table[x index={48}, y index={0}] {Figures/CDF_SE_rev1_052020.txt};

	\addplot [blue, mark=o, only marks,mark repeat=20,mark phase=0,mark options={solid, line width = 0.8pt},mark size=2pt]
	table[x index={62}, y index={0}] {Figures/CDF_SE_rev1_052020.txt};	
	
	\addplot [blue, mark=none,  loosely  dashed]
	table[x index={64}, y index={0}] {Figures/CDF_SE_rev1_052020.txt};


%
\addplot [green, mark=diamond*, only marks,mark repeat=20,mark phase=0,mark options={solid, line width = 0.8pt},mark size=2pt]
table[x index={23}, y index={0}] {Figures/CDF_SE_mxHigh1Qbit_tc200_062020.txt};

\addplot [green, mark=diamond, only marks,mark repeat=20,mark phase=0,mark options={solid, line width = 0.8pt},mark size=2pt]
table[x index={24}, y index={0}] {Figures/CDF_SE_mxHigh1Qbit_tc200_062020.txt};

	\end{axis}
	\end{tikzpicture}%

%% file: Figures/Pw_OP_GP_legend.tex

\begin{tikzpicture}
	\begin{axis}[
	width=0.8\textwidth,
	height=50pt,
	axis line style={draw=none},
	tick style={draw=none},
	xticklabels=\empty,
	yticklabels=\empty,
	xmin = 0,
	xmax = 500,
	ymin = 0,
	ymax = 1,
	line width = 1pt,
	legend columns = 4,
	legend cell align={left},
	legend style={at={(0.3,0)},anchor=north,font=\small,line width=1pt,draw=black,mark size=0.2pt},
	]

	\addlegendentry{\textit{Co-corr-I} mixed-ADC}
	\addlegendimage{black, mark=none, solid,mark options={solid, line width = 0.8pt},mark size=2pt}	
	\addlegendentry{\textit{Co-corr-D1} mixed-ADC}
	\addlegendimage{red, mark=none,    densely dashed,mark options={solid, line width = 0.8pt},mark size=2pt}
	\addlegendentry{\textit{Co-corr-DK} mixed-ADC}
	\addlegendimage{green, mark=none,    dashdotdotted,mark options={solid, line width = 0.8pt},mark size=2pt}
	\addlegendentry{\textit{Cell-free} mixed-ADC}
	\addlegendimage{blue, mark=none,    densely dotted,mark options={solid, line width = 0.8pt},mark size=2pt}

	\addlegendentry{\textit{Co-corr-I} eq. ADC}
	\addlegendimage{black, mark=square, only marks,mark options={solid, line width = 1.1pt},mark size=2.2pt}	
	\addlegendentry{\textit{Co-corr-D1} eq. ADC}
	\addlegendimage{red, mark=triangle,    only marks,mark options={solid, line width = 1.1pt},mark size=2.2pt}
	\addlegendentry{\textit{Co-corr-DK} eq. ADC}
	\addlegendimage{green, mark=x,    only marks,mark options={solid, line width = 1.1pt},mark size=2.2pt}
	\addlegendentry{\textit{Cell-free} eq. ADC}
	\addlegendimage{blue, mark=o,    only marks,mark options={solid, line width = 1.1pt},mark size=2.2pt}

	\end{axis}
	\end{tikzpicture}%

%% file: Figures/CDF_Qbit_maxSE_tgPw_GP_2020.tex

\begin{tikzpicture}
	\begin{axis}[
	width=0.47\textwidth,
	height=0.32\textwidth,
	y label style = {at={(-0.03,0.5)},anchor=north},
	xlabel={ADC bit resolution},
	ylabel={CDF},
	xmin = 0.9,
	xmax =6,
	ymin = 0,
	ymax = 1,
	line width = 1pt,
	grid=both,
	]	



\addplot [black, mark=none, solid]
table[x index={10}, y index={0}] {Figures/CDF_Qbit_tgPw_maxSE_GP_062020.txt};		
\addplot [black, mark=square, only marks,mark repeat=10,mark phase=1,mark options={solid, line width = 0.8pt},mark size=2pt]
table[x index={2}, y index={0}] {Figures/CDF_Qbit_tgPw_maxSE_GP_062020.txt};

\addplot [red, mark=triangle, only marks,mark repeat=10,mark phase=1,mark options={solid, line width = 0.8pt},mark size=2pt]
table[x index={12}, y index={0}] {Figures/CDF_Qbit_tgPw_maxSE_GP_062020.txt};

	\addplot [red, mark=none,    densely dashed]
	table[x index={4}, y index={0}] {Figures/CDF_Qbit_tgPw_maxSE_GP_062020.txt};

\addplot [green, mark=none,    dashdotdotted]
table[x index={6}, y index={0}] {Figures/CDF_Qbit_tgPw_maxSE_GP_062020.txt};	
	\addplot [green, mark=x, only marks,mark repeat=10,mark phase=1,mark options={solid, line width = 0.8pt},mark size=2pt]
	table[x index={14}, y index={0}] {Figures/CDF_Qbit_tgPw_maxSE_GP_062020.txt};

\addplot [blue, mark=none, densely dotted]
table[x index={8}, y index={0}] {Figures/CDF_Qbit_tgPw_maxSE_GP_062020.txt};		

\addplot [blue, mark=o,only marks,mark repeat=10,mark phase=1,mark options={solid, line width = 0.8pt},mark size=2pt]
table[x index={16}, y index={0}] {Figures/CDF_Qbit_tgPw_maxSE_GP_062020.txt};


%

	\end{axis}
	\end{tikzpicture}%



	

%% file: Figures/EE_vs_Pwtarget_GP.tex

\begin{tikzpicture}
	\begin{axis}[
	width=0.47\textwidth,
	height=0.32\textwidth,
	y label style = {at={(-0.03,0.5)},anchor=north},
	xlabel={Target power consumption $\gamma_\textsc{pc}$},
	ylabel={EE [Mbit/Joule]},
	xmin = 3,
	xmax = 30,
	ymin = 10,
	ymax = 41,
	xtick distance = 5,
	ytick distance = 10,
	line width = 1pt,
	grid=both,
	]	

\addplot [black, mark=none, solid]
table[x index={0}, y index={2}] {Figures/EEvstgPw_maxSE_GP.txt};						


%

\addplot [black, mark=square, only marks,,mark options={solid, line width = 0.8pt},mark size=1.9pt]
table[x index={0}, y index={18}] {Figures/EEvstgPw_maxSE_GP.txt};						

	\addplot [red, mark=none,    densely dashed]
	table[x index={0}, y index={6}] {Figures/EEvstgPw_maxSE_GP.txt};			


	\addplot [red, mark=triangle,     only marks,,mark options={solid, line width = 0.8pt},mark size=2.5pt]
	table[x index={0}, y index={22}] {Figures/EEvstgPw_maxSE_GP.txt};

\addplot [green, mark=none,dashdotdotted]
table[x index={0}, y index={10}] {Figures/EEvstgPw_maxSE_GP.txt};	 


\addplot [green, mark=x,only marks,,mark options={solid, line width = 0.8pt},mark size=2.8pt]
table[x index={0}, y index={26}] {Figures/EEvstgPw_maxSE_GP.txt};



	\addplot [blue, mark=none,     densely dotted]
	table[x index={0}, y index={14}] {Figures/EEvstgPw_maxSE_GP.txt};	

	\addplot [blue, mark=o,   only marks,,mark options={solid, line width = 0.8pt},mark size=2pt]
	table[x index={0}, y index={30}] {Figures/EEvstgPw_maxSE_GP.txt};	




	\end{axis}
	\end{tikzpicture}%
	